\documentclass[%
 reprint,
 amsmath,amssymb,
 prl,
]{revtex4-1}

\usepackage{graphicx}
\usepackage{dcolumn}
\usepackage{bm}

\usepackage{graphicx}
\usepackage{dcolumn}
\usepackage{lipsum}
\usepackage{mathtools}
\usepackage{amsmath}
\usepackage{bm}
\usepackage{float}
\usepackage{hyperref}
\usepackage{tabularx}
\usepackage{array}

\usepackage{tabularx}
\usepackage{array}
\newcolumntype{P}[1]{>{\centering\arraybackslash}p{#1}}
\usepackage{booktabs}

\begin{document}

\preprint{APS/123-QED}

\title{Entanglement without indistinguishability: Routing of coupling-induced path-entangled photon pairs}

\author{Raja Ahmad}
    \email{rahmad@ofsoptics.com}
    \altaffiliation[Present address: ]{OFS Laboratories, 19-Schoolhouse Road, Somerset, New Jersey 08873, USA}
    \affiliation{CREOL, The College of Optics \& Photonics, University of Central Florida, Orlando, Florida 32816, USA}
\author{Ayman F. Abouraddy}
    \email{raddy@creol.ucf.edu}
    \affiliation{CREOL, The College of Optics \& Photonics, University of Central Florida, Orlando, Florida 32816, USA}

\date{\today}

\begin{abstract}
Realizing an on-chip reconfigurable source of path-entangled photons is of critical importance for the advancement of quantum information processing and networking. Achieving this goal has proven challenging to date. We present an on-chip scheme for the deterministic creation of co-propagating or counter-propagating path-entangled photon pairs that can be routed in multiple configurations by tuning a classical parameter. The simplest manifestation of this approach makes use of two \textit{coupled} waveguides: a \textit{nonlinear} waveguide that produces photon pairs via spontaneous parametric downconversion from an externally incident unguided optical pump, and an auxiliary \textit{linear} waveguide. Although the photon pairs are born in only one waveguide, which alone cannot create path-entanglement, linear coupling over an extended length to the passive waveguide introduces unexpected indistinguishability that induces path-entanglement. Tuning of the classical pump spatial profile allows routing the photon pairs over all possible configurations. The proposed device is a building block for future quantum-optical networks.

\end{abstract}

\maketitle

Continued progress in the applications of quantum information science~\cite{NielsenQCQI10,GisinNP07,GiovannettiNP11} requires the reliable processing of entangled photons in systems of increasing complexity. By exploiting miniaturized versions of their free-space counterparts (e.g., linear couplers replacing beam splitters~\cite{BrombergPRL09,PeruzzoScience10,ShadboltNatP11,SilverstoneNPhoton14}), photonic integrated circuits outperform their free-space counterparts with regards to stability, scalability, and compactness~\cite{OBrienNP09} in implementations of two-photon interference, quantum logic operations, and quantum teleportation ~\cite{SilverstoneNPhoton14,PolitiScience08,OkamotoPNAS11,CrespiNComm11,BonneauNJPhys12,HWLiNJPhys11,MetcalfNP14}. Here, we explore a configuration for entanglement generation that is challenging to achieve in a free-space arrangement, but occurs naturally in an on-chip system enabled by extended evanescent coupling between parallel waveguides (WGs). 

Quantum photonic chips are usually \textit{passive} unitary transformations that can be viewed as a quantum walk~\cite{PeretsPRL08,RaiPRA08,BrombergPRL09,PeruzzoScience10,ShadboltNatP11,AbouraddyPRA12,DiGiuseppePRL13,GileadPRL15}. A salutary feature of this approach is the potential for combining processing of quantum states \textit{and} their active generation via spontaneous parametric downconversion (SPDC) in a single second-order WG~\cite{HornPRL12,OrieuxPRL13,BoitierPRL14} or four-wave mixing in a third-order nonlinear waveguide (WG)~\cite{MatsudaSCiRep12,SilverstoneNPhoton14}. Here, we demonstrate that \textit{extended} linear coupling between a \textit{nonlinear} WG (producing SPDC photons) and an auxiliary \textit{linear} WG (\textit{not} producing SPDC photons) provides a new mechanism for generating controllable two-photon path-entanglement. Traditionally, path-entanglement requires \textit{indistinguishability} between the pathways for photon-pair generation from multiple sources. In the configuration explored here, \textit{only one WG produces photons} and is thus clearly \textit{distinguishable} from the passive (linear) WG -- path-entanglement is nevertheless created. We refer to the physical principle underlying this scheme as `coupling-induced path-entanglement' (COPE), which results from a subtle interplay between \textit{phase-matching} in the nonlinear WG and \textit{linear coupling} along its length to an auxiliary linear WG.

Arrays of coupled linear~\cite{ChristodoulidesNat03,LedererPhysRep08} and nonlinear~\cite{ChristodoulidesNat03,LedererPhysRep08,KartashovRMP11,FlachPhysRep08,MalomedJOptB05} WGs have long been used in classical optics as a platform for studying complex dynamics. Recently, the propagation of a classical pump injected into an array of nonlinear WGs along with SPDC-generated photon pairs has been examined~\cite{SolntsevPRL12}. The linear walk of the pump and SPDC-photons in $N$ WGs modulates the pairwise correlations on an $N\!\times\!N$ combinatorial grid. Here, we show that tuning the parameters of a classical optical pump over \textit{one} nonlinear WG coupled to a linear WG facilitates reconfiguring photon-pair creation in an arbitrary configuration: entangled or separable, co-propagating or counter-propagating, correlated or anti-correlated ports -- thereby yielding a \textit{quantum router}. From a practical perspective, the coupled WGs will be of the same material in monolithic realizations, and pumping only one WG renders the other effectively linear. We provide a design for a periodically poled lithium-niobate (PPLN) device that can be used to explore on-chip routing of path-entangled photons.

\begin{figure}[t!]
\centering\includegraphics[width=8.6cm]{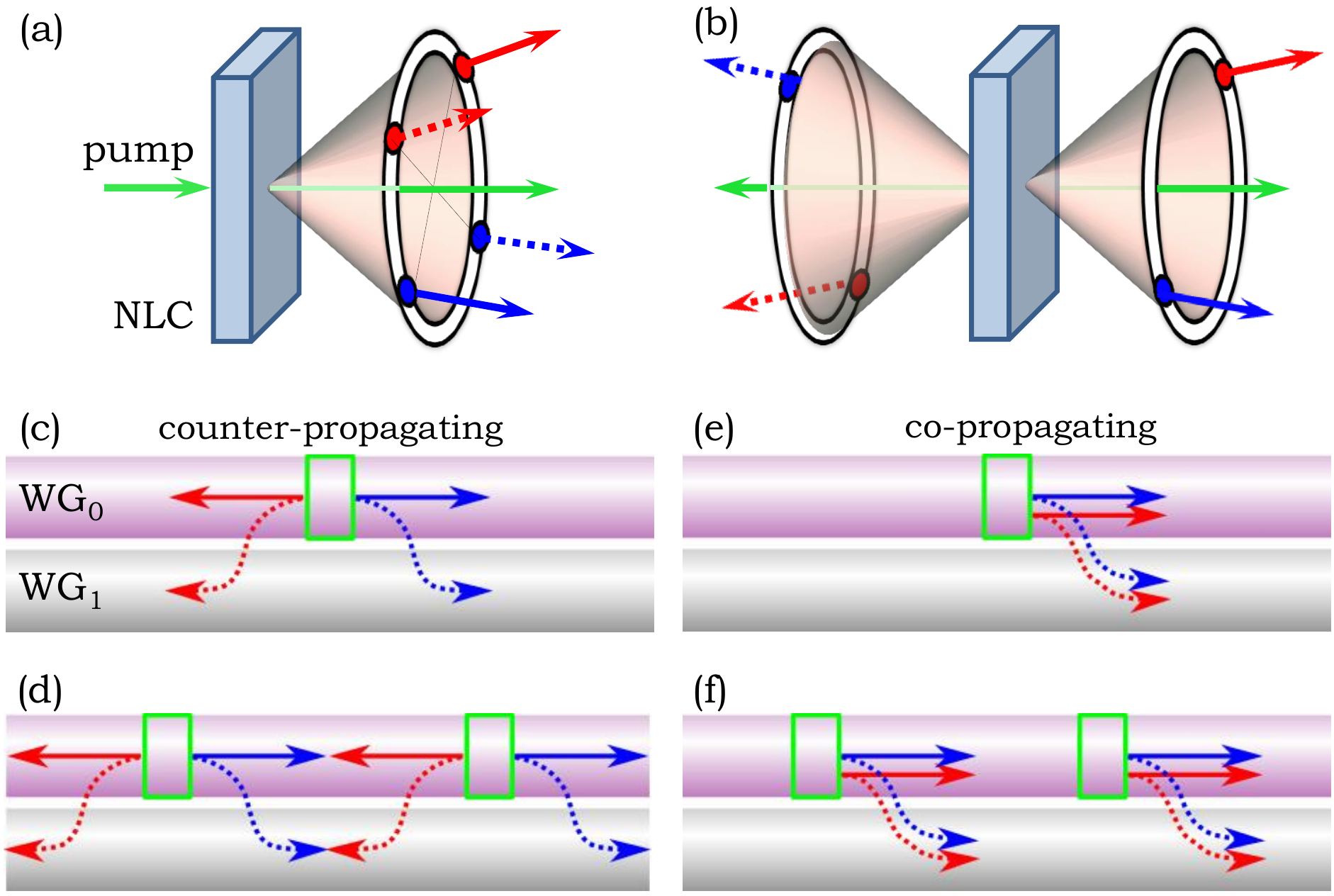}
\caption{Generation of path-entanglement from a two-photon source. (a),(b) Non-collinear SPDC producing photon pairs emitted in directions lying on a cone; NLC: nonlinear crystal. (c)-(f) The principle of coupling-induced path-entanglement (COPE). Only one of the two pairs shown in (d) and (f) is actually created, but the photons in that pair can now be path-entangled. The areas enclosed in green rectangles are the regions of photon-pair creation.}
\label{Fig1:General Idea}\vspace{-4mm}
\end{figure}

\textit{General Principle of COPE.} --- The first observation of path-entanglement made use of the arrangement in Fig.~\ref{Fig1:General Idea}(a), in which a monochromatic plane-wave pump incident on a nonlinear crystal produces pairs of photons via type-I SPDC~\cite{RarityPRL90,RarityPRA90}. Transverse momentum conservation produces exit angles for spectrally degenerate non-collinear photon pairs that are anti-correlated around a cone~\cite{SalehPRA00}. Selecting the four angles shown in Fig.~\ref{Fig1:General Idea}(a) selects a path-entangled state~\cite{RarityPRA90}. A variation on this theme for creating path-entanglement exploits a double-pass of a pump through the nonlinear crystal followed by a selection of two paths on opposing sides~\cite{PanPRL01,ChenPRL03}; Fig.~\ref{Fig1:General Idea}(b). In both cases, the multiple modes that define the indistinguishable entangled paths are a consequence of the three-dimensional nature of the nonlinear crystal. If instead the photons are emitted into well-defined spatial modes, as in a single-mode nonlinear WG, then path-entanglement \textit{cannot be subsequently introduced} via local unitary operations, such as beam splitters.


Merging these two aspects -- nonlinearity for photon generation and unitary transformations to introduce multiple paths -- into the \textit{same} structure can pave the way to generating reconfigurable path-entanglement. Our approach exploits two coupled WGs (WG$_0$ and WG$_1$), one of which (say, WG$_0$) is nonlinear; see Fig.~\ref{Fig1:General Idea}(c-f). Consider counter-propagating photons produced at position $z$ along WG$_0$ -- potentially by illuminating that position with an external pump laser; Fig.~\ref{Fig1:General Idea}(c). At the birth position, the two-photon state is separable and remains so at the exit of the WG system since each photon undergoes a local unitary transformation.

Now consider a scenario where the photon pairs are born at two symmetric positions with respect to the center of WG$_0$ with equal probability amplitudes and $CL\!=\!m\pi$, where $C$ is the coupling coefficient, $L$ is WG length, and $m$ is an integer; Fig.~\ref{Fig1:General Idea}(d). Quantum interference between the events of two-photon births eliminates the probability amplitudes of the photons emerging from different WGs, thus ensuring they always emerge either from WG$_0$ \textit{or} WG$_1$; i.e., in the path-entangled Bell-state $|\phi^{+}\rangle$. Alternatively, changing the coupling condition such that $CL\!=\!(2m+1)\frac{\pi}{2}$, eliminates the probability amplitudes of the photons emerging from the same WGs, thus ensuring they always emerge from opposing WGs; i.e., in the path-entangled Bell-state $|\psi^{+}\rangle$. In both scenarios, it is the linear coupling between the WGs extending over the same region of potential photon-birth in only \textit{one} nonlinear WG that creates path-entanglement. Similar arguments apply to a configuration in which the photon pair co-propagate in the same direction; Fig.~\ref{Fig1:General Idea}(e-f). We extend this heuristic description below to the more realistic configuration of photon pairs generated over an extended section of WG$_0$. We refer to this phenomenon as `COupling-induced Path Entanglement' (COPE).

\begin{figure}[t!]
\begin{center}
\includegraphics[width=8.6cm]{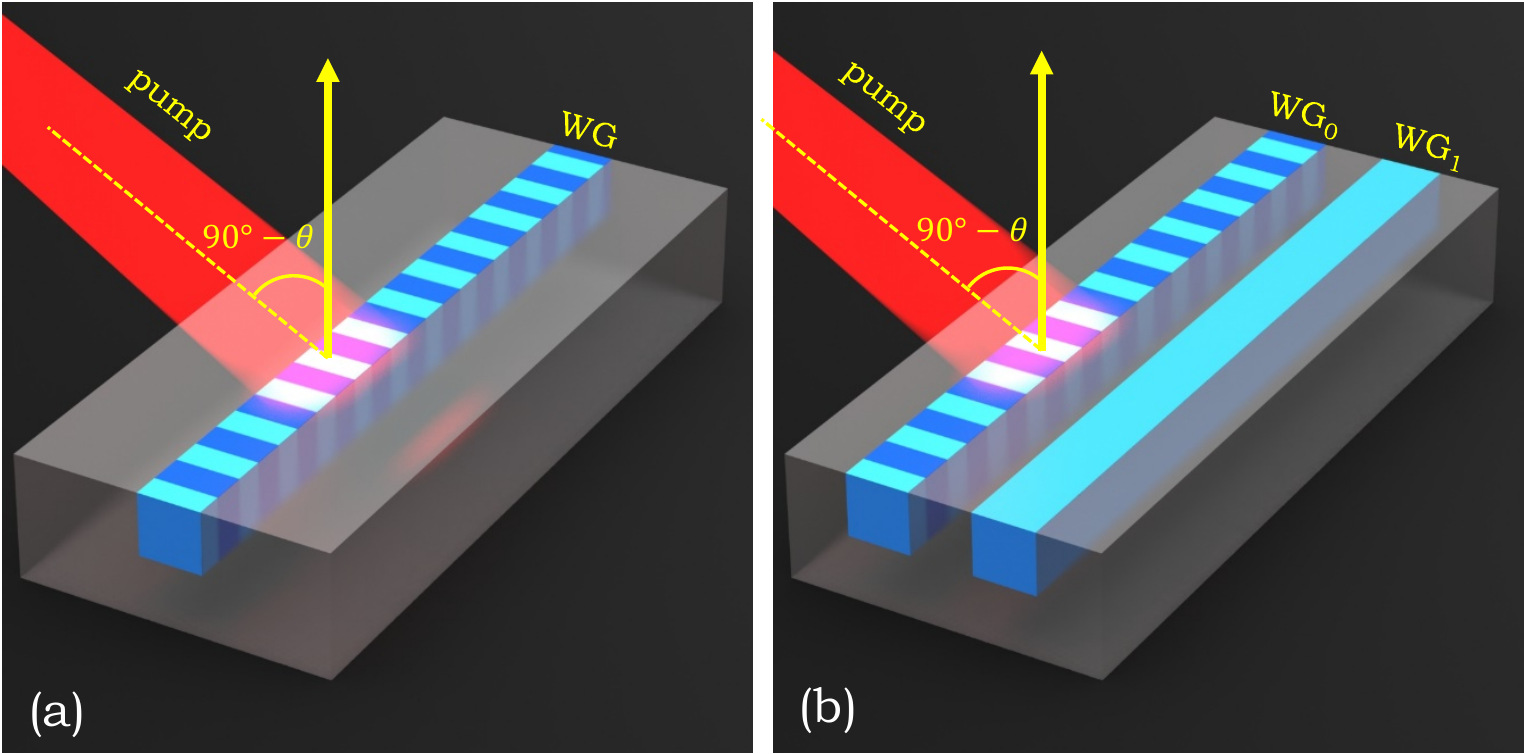}
    \label{Fig2:PMC}
\end{center}\vspace{-7mm}
\caption{(a) Schematic of a PPLN WG pumped by an obliquely incident, unguided classical pump beam for producing photon pairs via SPDC. (b) Schematic of a pair of evanescently coupled WGs, one of which (WG$_0$) is nonlinear and identical to the PPLN WG in (a), while the other (WG$_1$) is linear. The pump illuminates an extended section of the nonlinear WG and is incident at an angle $90^{\circ}\!\!-\!\theta$ with respect to the normal.}
\label{Fig2:Schematic and Modes}\vspace{-3mm}
\end{figure}

The principle described here is antithesis to the accepted wisdom whereupon the superposition of \textit{indistinguishable} events is required to create entanglement. In general, indistinguishability is critical to a host of quantum optical effects such as two-photon interference~\cite{HongPRL87} and also plays a crucial role in Mandel's well-known experiment~\cite{ZouPRL91,LemosNat14} whereupon the indistinguishability of the idler photons produced from two different sources induces coherence between the associated signal photons. In contradistinction, the photons in the COPE scheme are created in WG$_0$ alone, and never in WG$_1$. Nevertheless, this fundamental distinguishability of the photon source is wiped out by the spatially extended coupling.

\textit{Photon pairs generated in a nonlinear WG.} --- We make use of the model for WG SPDC in Ref.~\cite{BoothPRA02}, but we emphasize that any alternative model can serve as a building block. The quantum state associated with SPDC photon pairs produced in one WG is
\begin{equation}\label{Eq1:Simplified state equation}
|\Psi\rangle=\int\!\!\!\int\!d\omega_{\mathrm{s}}d\omega_{\mathrm{i}}\, \Phi(\omega_{\mathrm{s}},\omega_{\mathrm{i}})|1_{\mathrm{s}},1_{\mathrm{i}}\rangle,
\end{equation}
where $\int\!\!\int \!\!d\omega_{\mathrm{s}}d\omega_{\mathrm{i}}|\Phi(\omega_{\mathrm{s}},\omega_{\mathrm{i}})|^{2}\!=\!1$, and $\Phi(\omega_{\mathrm{s}},\omega_{\mathrm{i}})$ is determined by phase-matching and describes the correlations between the signal and idler frequencies $\omega_{\mathrm{s}}$ and $\omega_{\mathrm{i}}$, respectively. Each photon is emitted into the \textit{same} WG spatial mode $u(\textbf{r},\omega)$, where \textbf{r} is the transverse spatial vector ~\cite{BoothPRA02}. We assume a monochromatic pump at a frequency $\omega_{\mathrm{p}}$, such that $\omega_{\mathrm{s}}\!+\!\omega_{\mathrm{i}}\!=\!\omega_{\mathrm{p}}$, obliquely incident on the WG [Fig.~\ref{Fig2:Schematic and Modes}(a)] whose profile $E_{\mathrm{p}}(\mathbf{r},z)\!\approx\!E_{\mathrm{p}}(z)$ varies along the WG ($z$ coordinate) but not appreciably in the WG cross section. It can be shown that
\begin{equation}\label{Eq2:State function general}
\Phi(\omega_{s},\omega_{i})\!\!\propto\!\!\int\!\!\!\!\int\!d^{2}\textbf{r}\,u(\textbf{r},\omega_{\mathrm{s}}) u(\textbf{r},\omega_{\mathrm{i}})\cdot\!\!\int_{0}^{L}\!\!\!\!d{z}E_{\mathrm{p}}(z)e^{\imath\Delta\beta_{\ell}z},
\end{equation}
where $\Delta\beta_{\ell}\!\!=\!\!\beta_{\mathrm{p}}\!\!-\!\!\beta_{\mathrm{s}}\!\!-\!\!\beta_{\mathrm{i}}\!-\!K_{\ell}$ is the axial-wave-number mismatch, $K_{\ell}\!=\!2\pi\ell\!/\!\Lambda$, $\Lambda$ is the poling period in an appropriately designed PPLN crystal, $\ell$ is the Fourier-order of the spatially poled WG, and $\beta_{\mathrm{s}}$, $\beta_{\mathrm{i}}$, and $\beta_{\mathrm{p}}$ are the axial wave numbers for the signal, idler, and pump, respectively. The sign convention we adopt is that axial wave numbers have the same (opposite) sign for co-propagating (counter-propagating) photons. By tuning the pump incidence angle $\theta$, and hence the pump axial wave vector $\beta_{\mathrm{p}}\!=\!k_{\mathrm{p}}\cos\theta$, the phase-matching condition ($\Delta \beta_{\ell}$=0) can be satisfied for either \emph{counter}-propagating or \emph{co}-propagating photons~\cite{DeRossiPRL02,BoothPRA02}. For subsequent use, we divide the WG length $L$ into $N$ segments of length $\Delta L\!=\!L\!/\!N$ with piecewise constant pump amplitude $E_{n}$ in the $n^{\mathrm{th}}$ section. Here $\Phi$ becomes a superposition of probability amplitudes associated with these segments,
\begin{equation}\label{Eq2b:State function with segmented WG}
\Phi(\omega_{s},\omega_{i})\,\,\propto\,\,\mathrm{sinc}\left(\frac{\varphi}{\pi}\right)\sum\limits_{n=0}^{N-1} e^{\imath\varphi(2n+1)}E_{n}=f(\varphi),
\end{equation}
where $\varphi\!=\!\Delta\beta_{\ell}\Delta L\!/\!2$.

\textit{Path-entangled counter-propagating photons from a pair of coupled WGs}\label{Counter-prop_entangled_photons.} --- Consider a system comprising two parallel linearly coupled WGs (WG$_0$ is nonlinear and WG$_1$ is linear) with coupling constants $C_{\mathrm{s}}$ and $C_{\mathrm{i}}$ at the signal and idler wavelengths, respectively; see Fig.~\ref{Fig2:Schematic and Modes}(b). We assume that $C$ varies linearly with wavelength~\cite{BortzOL91,LedererPhysRep08}. A localized pump at $z$ on WG$_0$ at a suitable incidence angle leads to the birth of two \textit{counter}-propagating photons from $z$. The signal photon propagating to the right travels through a linear coupler of length $z$, while the idler propagates to the left through a linear coupler of length $L\!-\!z$. The signal and idler evolve according to $\hat{a}_{\mathrm{s}}^{\dagger}|0\rangle\!\rightarrow\!\alpha_{\mathrm{s}}|\mathrm{s}_{0}\rangle\!+\!\beta_{\mathrm{s}}|\mathrm{s}_{1}\rangle$ and $\hat{a}_{\mathrm{i}}^{\dagger}|0\rangle\!\rightarrow\!\alpha_{\mathrm{i}}|\mathrm{i}_{0}\rangle\!+\!\beta_{\mathrm{i}}|\mathrm{i}_{1}\rangle$, respectively, resulting in a path-separable two-photon state
\begin{equation}\label{SeparableState_PointExcitation}
|\Psi^{(0)}\!(z)\rangle=\left\{\alpha_{\mathrm{s}}|\mathrm{s}_{0}\rangle\!+\!\beta_{\mathrm{s}}|\mathrm{s}_{1}\rangle\right\}\!\otimes\!\left\{\alpha_{\mathrm{i}}|\mathrm{i}_{0}\rangle\!+\!\beta_{\mathrm{i}}|\mathrm{i}_{1}\rangle\right\},
\end{equation}
where the state coefficients are $\alpha_{\mathrm{s}}\!=\!\cos{C_{s}(L\!-\!z)}$, $\beta_{\mathrm{s}}\!=\!-\imath\sin{C_s(L\!-\!z)}$, $\alpha_{\mathrm{i}}\!=\!\cos{C_{i}z}$, and $\beta_{\mathrm{i}}\!=\!-\imath\sin{C_{i}z}$. The signal and idler photons have finite probabilities of being detected in WG$_0$ \textit{and} WG$_1$, but the state is of course \textit{separable}; a unitary transformation can undo the effect of linear coupling, converting it back to $|\mathrm{s}_{0}\rangle\!\otimes\!|\mathrm{i}_{0}\rangle$.

Increasing the spatial extent of the pump illumination $E_{\mathrm{p}}(z)$ to the entire WG$_0$ produces the path-entangled two-photon state
\begin{equation}\label{Eq7:State function two pumps}
|\Psi\rangle\!=\!\!\int\!\!\!\!\int \!\!d\omega_{\mathrm{s}}d\omega_{\mathrm{i}}\!\!\int\!\!\!\!\int\!\!d{^{2}\mathbf{r}}\,u(\mathbf{r},\omega_{\mathrm{s}})u(\mathbf{r},\omega_{\mathrm{i}})\cdot\!\!\int_{0}^{L}\!\!\!d{z}\,E_{\mathrm{p}}\!(z)e^{\imath\Delta\beta_{\ell} z}|\Psi^{(0)}\!(z)\rangle.
\end{equation}
We extract from Eq.~\ref{Eq7:State function two pumps} the two-photon probability amplitudes $M_{mm'}$ for the arrival of a signal photon on the right in WG$_m$ and an idler photon on the left in WG$_{m'}$, $m,m'\!=\!0,1$, which are normalized such that $|M_{00}|^{2}\!+\!|M_{01}|^{2}\!+\!|M_{10}|^{2}\!+\!|M_{11}|^{2}\!=\!1$. These probability amplitudes determine the degree of path-entanglement $D_{\mathrm{p}}\!=\!2|M_{00}M_{11}\!-\!M_{01}M_{10}|$~\cite{AbouraddyPRA01}. As in the case of the single nonlinear WG, we segment the length $L$ of WG$_0$ into $N$ sections of length $\Delta L\!=\!L\!/\!N$ each with piecewise constant pump amplitudes. The four probability amplitudes $M_{mm'}$ for the counter-propagating photons can be cast in the general form:
\begin{equation}\label{Eq4:M_values_counter_prop}
\left[\!\begin{array}{c}\mathrm{M}_{00} \\ \mathrm{M}_{01} \\ \mathrm{M}_{10} \\ \mathrm{M}_{11} \end{array}\!\right]
\!\!\propto\!
\left[\begin{array}{cccc}
        +1 & +1 & +1 & +1 \\
        -1 & -1 & +1 & +1 \\
        -1 & +1 & -1 & +1 \\
        +1 & -1 & -1 & +1 \\
       \end{array}\right]\!
\left[\!\!\begin{array}{c}
e^{\imath C_{s}L}f(\varphi_{-+})\\
e^{-\imath C_{s}L}f(\varphi_{++})\\
e^{\imath C_{s}L}f(\varphi_{--})\\
e^{-\imath C_{s}L}f(\varphi_{+-})
     \end{array}\!\!\right]\!.
\end{equation}
The function $f(\varphi)$ is defined in Eq.~\ref{Eq2b:State function with segmented WG} and the phases are given by $\varphi_{-+}\!=\!(\Delta\beta_{\ell}\!-\!C_{\mathrm{s}}\!+\!C_{\mathrm{i}})\Delta L\!/\!2$, $\varphi_{++}\!=\!(\Delta\beta_{\ell}\!+\!C_{\mathrm{s}}\!+\!C_{\mathrm{i}})\Delta L\!/\!2$, $\varphi_{--}\!=\!(\Delta\beta_{\ell}\!-\!C_{\mathrm{s}}\!-\!C_{\mathrm{i}})\Delta L\!/\!2$, and $\varphi_{+-}\!=\!(\Delta\beta_{\ell}\!+\!C_{\mathrm{s}}\!-\!C_{\mathrm{i}})\Delta L\!/\!2$.

Reducing any of the phases ($\varphi_{-+}$, $\varphi_{++}$, $\varphi_{--}$, or $\varphi_{+-}$) to 0 is achieved by selecting the appropriate pump incidence angle $\theta$, for which we use the same notation: $\theta_{-+}$ is the angle that sets $\varphi_{-+}\!=\!0$, etc. Selecting an incidence angle $\theta$ amounts to providing the pump wave front with a linear phase, $E_{\mathrm{p}}(z)\!=\!E_{\mathrm{o}}e^{ik_{\mathrm{p}}z\cos{\theta}}$, where $E_{\mathrm{o}}$ is a constant if we take all the $E_{n}$'s to be equal. This linear phase is readily provided by a spatial light modulator. Multiple incidence angles may be excited simultaneously by judiciously modulating the pump wave fronts; for example, preparing the wave front $E_{\mathrm{p}}(z)\!=\!E_{\mathrm{o}}\{a_{1}e^{ik_{\mathrm{p}}z\cos{\theta_{++}}}\!+\!a_{2}e^{ik_{\mathrm{p}}z\cos{\theta_{+-}}}\}$ satisfies the phase-matching conditions $\varphi_{++}\!=\!\varphi_{+-}\!=\!0$, simultaneously; $a_{1}$ and $a_{2}$ are the relative excitation weights.  

The equations derived above predict the possibility of generating both frequency- and path-entanglement between the two photons. For simplicity, we consider hereafter spectrally degenerate photons ($\omega_{\mathrm{s}}\!=\!\omega_{\mathrm{i}}$, hence $C_{\mathrm{s}}\!=\!C_{\mathrm{i}}\!=\!C$) to focus on path-entanglement. Spectral degeneracy reduces the phases to $\varphi_{-+}\!=\!\varphi_{+-}\!=\!\Delta\beta_{\ell}\Delta L\!/\!2$, $\varphi_{++}\!=\!(\!\Delta\beta_{\ell}\!+\!2C)\Delta L\!/\!2$, and $\varphi_{--}\!=\!(\Delta\beta_{\ell}\!-\!2C)\Delta L\!/\!2$. In this special case, $\theta_{-+}\!=\!\theta_{+-}\!=\!\theta_{0}$. Control over path-entanglement in this configuration can be exercised by sculpting the pump profile along WG$_0$ or tuning the coupling coefficient $C$ in a planar WG platform by electro-optic, thermal, or other means~\cite{GuarinoNPhoton07,RabeloAO11,LocatelliOE12}. Consequently, path-entanglement is readily \textit{created} and \textit{reconfigured}, such that the photon pair is \textit{routed} to the desired WGs in any linear combination of Bell states.

By selecting the pump incidence angle $\theta_{0}$, such that $E_{\mathrm{p}}(z)\!=\!E_{\mathrm{o}}e^{ik_{\mathrm{p}}z\cos{\theta_{0}}}$ and $\varphi_{-+}\!=\!\varphi_{+-}\!=\!0$, the counter-propagating two-photon state is
\begin{equation} \label{Eq5:Counter_prop_discrete_angle}
|\Psi\rangle=\cos{CL}|\phi^{+}\rangle-\imath\sin{CL}|\psi^{+}\rangle.
\end{equation}
Alternatively, by preparing two pump beams incident at $\theta_{++}$ and $\theta_{--}$, $E_{\mathrm{p}}(z)\!=\!E_{\mathrm{o}}\{e^{ik_{\mathrm{p}}z\cos{\theta_{++}}}\!+\!e^{ik_{\mathrm{p}}z\cos{\theta_{--}}}\}$, which helps satisfy $\varphi_{++}\!=\!\varphi_{--}\!=\!0$, produces the state
\begin{equation} \label{Eq6:Counter_prop_discrete_angle}
|\Psi\rangle=\cos{CL}\;|\phi^{-}\rangle+\imath\sin{CL}\;|\psi^{-}\rangle.
\end{equation}
Here, $|\phi^{\pm}\!\rangle\!=\!\!\tfrac{1}{\sqrt{2}}\!(|00\rangle\pm|11\rangle)$ and $|\psi^{\pm}\rangle\!=\!\!\tfrac{1}{\sqrt{2}}\!(|01\rangle\pm|10\rangle)$, where $|01\rangle$ denotes the signal photon emerging from WG$_0$ and the idler from WG$_1$, and so on. The states $|\phi^{\pm}\rangle$ correspond to the two photons always emerging from the same WGs; whereas the states $|\psi^{\pm}\rangle$ correspond to the two photons always emerging from different WGs. Therefore, with a proper choice of coupling coefficient $C$ and pump profile $E_{\mathrm{p}}(z)$, path-entanglement can be switched continuously from one Bell state to another.

The photons can be routed to the \textit{same WG} by setting $CL\!=\!m\pi$ and the pump incidence angle to $\theta_{0}$, thus producing the \textit{path-entangled} Bell state $|\phi^{+}\rangle$ in which the photons emerge always from the same WG. Alternatively, the Bell state $|\phi^{-}\rangle$ can be produced by utilizing a pump incident at the angles $\theta_{++}$ and $\theta_{--}$. Producing a \textit{separable} state in which the two photons are both routed to WG$_0$, $|\Psi\rangle\!=\!|00\rangle$, a pump incident at the three angles $\theta_{0}$, $\theta_{++}$, and $\theta_{--}$ is required soo that $E_{\mathrm{p}}(z)\!\propto\!e^{ik_{\mathrm{p}}z\cos{\theta_{0}}}\!+\!e^{ik_{\mathrm{p}}z\cos{\theta_{++}}}\!+\!e^{ik_{\mathrm{p}}z\cos{\theta_{--}}}$. To produce the remaining separable state $|\Psi\rangle\!=\!|11\rangle$, a pump profile $E_{\mathrm{p}}(z)\!\propto\!e^{ik_{\mathrm{p}}z\cos{\theta_{0}}}\!-\!e^{ik_{\mathrm{p}}z\cos{\theta_{++}}}\!-\!e^{ik_{\mathrm{p}}z\cos{\theta_{--}}}$ is needed.

Alternatively, routing the photons to \textit{different} WGs requires $CL\!=\!(m\!+\!\frac{1}{2})\pi$ and selecting the pump incidence angle $\theta_{0}$ to produce the Bell state $|\psi^{+}\rangle$, in which the path-entangled photons always emerge from opposing WGs. Furthermore, exploiting the configuration in which the pump is incident at the angles $\theta_{++}$ and $\theta_{--}$, the Bell state $|\psi^{-}\rangle$ is produced in which the photons also emerge from the same WG. To produce separable states in which the two photons are routed to different WGs, namely $|01\rangle$ and $|10\rangle$, requires combining both these configurations; that is, assigning pumps at the three angles $\theta_{0}$, $\theta_{++}$, and $\theta_{--}$, similarly to the previous subsection. The conditions for routing counter-propagating photons in different WGs are summarized in Table~\ref{tab:conditions map}.

\begin{table}[t!]
\centering
\caption{\textbf{Summary of the conditions for routing counter-propagating path-entangled photon pairs towards the desired WG ports.} The WG enclosed in a green boundary is the nonlinear source of the photon pairs; blue for signal, red for idler. The arrows indicate the photon paths. Solid and dotted arrows denote superposed events, and thus a path-entangled state. $\theta_{\mathrm{p}}$ is the pump incidence angle required to produce the two-photon state.}
\begin{tabular}{P{1cm} P{3.0cm} P{1.5cm} P{2.5cm}}
\hline
\specialrule{0.01pt}{0.01pt}{0.01pt}
State & Configuration & \multicolumn{2}{c}{\hspace{-8mm} Requirements} \\[0mm]
& & $CL$ & $\theta_{p}$ \\
\specialrule{0.01pt}{0.01pt}{0.01pt}
\\[-2mm]
$|00\rangle$ & \raisebox{-0.5\totalheight} {\includegraphics[width=0.12\textwidth,keepaspectratio]{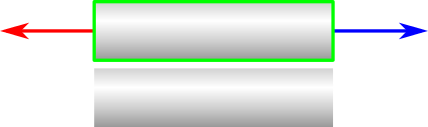}} & $m\pi$ & $\theta_{0}$, $\theta_{+-}$, $\theta_{++}$ \\[3mm]

$|11\rangle$ & \raisebox{-0.5\totalheight} {\includegraphics[width=0.12\textwidth,keepaspectratio]{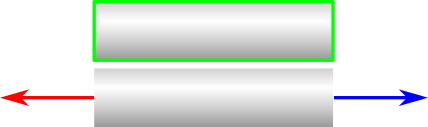}} & $m\pi$ & $\theta_{0}$, $\theta_{+-}$, $\theta_{++}$  \\[3mm]

$|\phi^{+}\rangle$ & \raisebox{-0.5\totalheight} {\includegraphics[width=0.12\textwidth,keepaspectratio]{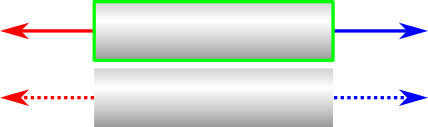}} & $m\pi$ & $\theta_{0}$ \\[3mm]

$|\phi^{-}\rangle$ & \raisebox{-0.5\totalheight} {\includegraphics[width=0.12\textwidth,keepaspectratio]{b5.png}} & $m\pi$ & $\theta_{+-}$, $\theta_{++}$  \\[7mm]

$|01\rangle$ & \raisebox{-0.5\totalheight} {\includegraphics[width=0.12\textwidth,keepaspectratio]{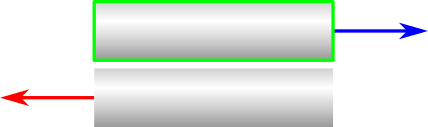}} & $(m\!+\!\!\tfrac{1}{2})\pi$ & $\theta_{0}$, $\theta_{+-}$, $\theta_{++}$ \\[3mm]

$|10\rangle$ & \raisebox{-0.5\totalheight} {\includegraphics[width=0.12\textwidth,keepaspectratio]{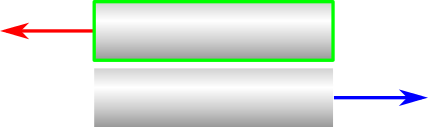}} & $(m\!+\!\!\frac{1}{2})\pi$ & $\theta_{0}$, $\theta_{+-}$, $\theta_{++}$ \\[3mm]

$|\psi^{+}\rangle$ & \raisebox{-0.5\totalheight} {\includegraphics[width=0.12\textwidth,keepaspectratio]{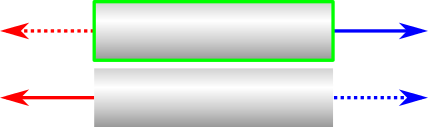}} & $(m\!+\!\!\frac{1}{2})\pi$ & $\theta_{0}$ \\[3mm]

$|\psi^{-}\rangle$ & \raisebox{-0.5\totalheight} {\includegraphics[width=0.12\textwidth,keepaspectratio]{b6.png}} & $(m\!+\!\!\frac{1}{2})\pi$ & $\theta_{+-}$, $\theta_{++}$ \\[4mm]
\hline
\end{tabular}
  \label{tab:conditions map}\vspace{-5mm}
\end{table}

We have explored the simplest case of linear coupling between two WGs. Extending our work to systems of three or more coupled WGs opens up the possibility of routing path-entangled photons across a large network through judicious modulation of the classical pump profile. The feasibility of implementing our proposed system in an integrated photonic-chip is demonstrated by a detailed design of a PPLN-based system consisting of two evanescently-coupled waveguides, one of which is pumped to generated photon pairs and the other acts as a passive waveguide. The design suggests the potential for implementing this concept with current technology \cite{SI}. Furthermore, other platform choices besides PPLN can be utilized, such as InP~\cite{Shi_OL_08_InP}, GaN~\cite{JLOBrien_APL_11_GaN_coupler} and InGaAsP~\cite{Griffel_PTL_00_InGaAsP_coupler}, which offer higher coupling coefficients and may be suitable candidates for compact quantum-routing devices. An intriguing possibility is a lithium-niobate-on-silicon platform~\cite{RabieiOE13} to monolithically generate photon pairs in lithium-niobate WGs and fabricate photodetectors on silicon. Finally, this scheme can be realized in other coupled components, including microresonators~\cite{GorodetskyJOSAB99,CaiPRL00,ArmaniNat03} and optical fibers~\cite{SumetskyOE04,PollingerPRL09,SumetskyOL10}.

In conclusion, we have proposed a technique for the deterministic creation and reconfiguration of path-entangled photon pairs in two linearly coupled WGs, one of which is nonlinear. Path-entanglement is enabled by the evanescent coupling between the modes of neighboring WGs. This is quite remarkable, as it allows the path-entanglement with one of the coupled WGs being \textit{passive} and not contributing to the SPDC process. That is, the WG in which the photon pair is born is completely distinguishable -- yet the paths encompassing both WGs become indistinguishable. The underlying physics of this novel entanglement-creating arrangement is that the coupling of two propagating photons during generation on the same platform introduces path-indistinguishability, and hence creates entanglement. The pump spatial profile controlled dynamically using spatial light modulators can route path-entangled or path-separable photons. Such states generated on a chip can be useful in applications ranging from imaging and microscopy, to routing quantum information in a network.

This work was supported by the U.S. Air Force Office of Scientific Research (AFOSR) MURI contract FA9550-14-1-0037. We thank F. Tan, A. K. Jahromi, and H. E. Kondakci for assistance with figure preparation.

\bibliography{References2}

\begin{thebibliography}{55}%
\makeatletter
\providecommand \@ifxundefined [1]{%
 \@ifx{#1\undefined}
}%
\providecommand \@ifnum [1]{%
 \ifnum #1\expandafter \@firstoftwo
 \else \expandafter \@secondoftwo
 \fi
}%
\providecommand \@ifx [1]{%
 \ifx #1\expandafter \@firstoftwo
 \else \expandafter \@secondoftwo
 \fi
}%
\providecommand \natexlab [1]{#1}%
\providecommand \enquote  [1]{``#1''}%
\providecommand \bibnamefont  [1]{#1}%
\providecommand \bibfnamefont [1]{#1}%
\providecommand \citenamefont [1]{#1}%
\providecommand \href@noop [0]{\@secondoftwo}%
\providecommand \href [0]{\begingroup \@sanitize@url \@href}%
\providecommand \@href[1]{\@@startlink{#1}\@@href}%
\providecommand \@@href[1]{\endgroup#1\@@endlink}%
\providecommand \@sanitize@url [0]{\catcode `\\12\catcode `\$12\catcode
  `\&12\catcode `\#12\catcode `\^12\catcode `\_12\catcode `\%12\relax}%
\providecommand \@@startlink[1]{}%
\providecommand \@@endlink[0]{}%
\providecommand \url  [0]{\begingroup\@sanitize@url \@url }%
\providecommand \@url [1]{\endgroup\@href {#1}{\urlprefix }}%
\providecommand \urlprefix  [0]{URL }%
\providecommand \Eprint [0]{\href }%
\providecommand \doibase [0]{http://dx.doi.org/}%
\providecommand \selectlanguage [0]{\@gobble}%
\providecommand \bibinfo  [0]{\@secondoftwo}%
\providecommand \bibfield  [0]{\@secondoftwo}%
\providecommand \translation [1]{[#1]}%
\providecommand \BibitemOpen [0]{}%
\providecommand \bibitemStop [0]{}%
\providecommand \bibitemNoStop [0]{.\EOS\space}%
\providecommand \EOS [0]{\spacefactor3000\relax}%
\providecommand \BibitemShut  [1]{\csname bibitem#1\endcsname}%
\let\auto@bib@innerbib\@empty
\bibitem [{\citenamefont {Nielsen}\ and\ \citenamefont
  {Chuang}(2010)}]{NielsenQCQI10}%
  \BibitemOpen
  \bibfield  {author} {\bibinfo {author} {\bibfnamefont {M.~A.}\ \bibnamefont
  {Nielsen}}\ and\ \bibinfo {author} {\bibfnamefont {I.~L.}\ \bibnamefont
  {Chuang}},\ }\href@noop {} {\emph {\bibinfo {title} {Quantum computation and
  quantum information}}}\ (\bibinfo  {publisher} {Cambridge university press},\
  \bibinfo {year} {2010})\BibitemShut {NoStop}%
\bibitem [{\citenamefont {Gisin}\ and\ \citenamefont {Thew}(2007)}]{GisinNP07}%
  \BibitemOpen
  \bibfield  {author} {\bibinfo {author} {\bibfnamefont {N.}~\bibnamefont
  {Gisin}}\ and\ \bibinfo {author} {\bibfnamefont {R.}~\bibnamefont {Thew}},\
  }\href@noop {} {\bibfield  {journal} {\bibinfo  {journal} {Nat. Photon.}\
  }\textbf {\bibinfo {volume} {1}},\ \bibinfo {pages} {165} (\bibinfo {year}
  {2007})}\BibitemShut {NoStop}%
\bibitem [{\citenamefont {Giovannetti}\ \emph {et~al.}(2011)\citenamefont
  {Giovannetti}, \citenamefont {Lloyd},\ and\ \citenamefont
  {Maccone}}]{GiovannettiNP11}%
  \BibitemOpen
  \bibfield  {author} {\bibinfo {author} {\bibfnamefont {V.}~\bibnamefont
  {Giovannetti}}, \bibinfo {author} {\bibfnamefont {S.}~\bibnamefont {Lloyd}},
  \ and\ \bibinfo {author} {\bibfnamefont {L.}~\bibnamefont {Maccone}},\
  }\href@noop {} {\bibfield  {journal} {\bibinfo  {journal} {Nat. Photon.}\
  }\textbf {\bibinfo {volume} {5}},\ \bibinfo {pages} {222} (\bibinfo {year}
  {2011})}\BibitemShut {NoStop}%
\bibitem [{\citenamefont {Bromberg}\ \emph {et~al.}(2009)\citenamefont
  {Bromberg}, \citenamefont {Lahini}, \citenamefont {Morandotti},\ and\
  \citenamefont {Silberberg}}]{BrombergPRL09}%
  \BibitemOpen
  \bibfield  {author} {\bibinfo {author} {\bibfnamefont {Y.}~\bibnamefont
  {Bromberg}}, \bibinfo {author} {\bibfnamefont {Y.}~\bibnamefont {Lahini}},
  \bibinfo {author} {\bibfnamefont {R.}~\bibnamefont {Morandotti}}, \ and\
  \bibinfo {author} {\bibfnamefont {Y.}~\bibnamefont {Silberberg}},\
  }\href@noop {} {\bibfield  {journal} {\bibinfo  {journal} {Phys. Rev. Lett.}\
  }\textbf {\bibinfo {volume} {102}},\ \bibinfo {pages} {253904} (\bibinfo
  {year} {2009})}\BibitemShut {NoStop}%
\bibitem [{\citenamefont {Peruzzo}\ \emph {et~al.}(2010)\citenamefont
  {Peruzzo}, \citenamefont {Lobino}, \citenamefont {Matthews}, \citenamefont
  {Matsuda}, \citenamefont {Politi}, \citenamefont {Poulios}, \citenamefont
  {Zhou}, \citenamefont {Lahini}, \citenamefont {Ismail}, \citenamefont
  {W{\"o}rhoff}, \citenamefont {Bromberg}, \citenamefont {Silberberg},
  \citenamefont {Thompson},\ and\ \citenamefont {O'Brien}}]{PeruzzoScience10}%
  \BibitemOpen
  \bibfield  {author} {\bibinfo {author} {\bibfnamefont {A.}~\bibnamefont
  {Peruzzo}}, \bibinfo {author} {\bibfnamefont {M.}~\bibnamefont {Lobino}},
  \bibinfo {author} {\bibfnamefont {J.~C.~F.}\ \bibnamefont {Matthews}},
  \bibinfo {author} {\bibfnamefont {N.}~\bibnamefont {Matsuda}}, \bibinfo
  {author} {\bibfnamefont {A.}~\bibnamefont {Politi}}, \bibinfo {author}
  {\bibfnamefont {K.}~\bibnamefont {Poulios}}, \bibinfo {author} {\bibfnamefont
  {X.-Q.}\ \bibnamefont {Zhou}}, \bibinfo {author} {\bibfnamefont
  {Y.}~\bibnamefont {Lahini}}, \bibinfo {author} {\bibfnamefont
  {N.}~\bibnamefont {Ismail}}, \bibinfo {author} {\bibfnamefont
  {K.}~\bibnamefont {W{\"o}rhoff}}, \bibinfo {author} {\bibfnamefont
  {Y.}~\bibnamefont {Bromberg}}, \bibinfo {author} {\bibfnamefont
  {Y.}~\bibnamefont {Silberberg}}, \bibinfo {author} {\bibfnamefont {M.~G.}\
  \bibnamefont {Thompson}}, \ and\ \bibinfo {author} {\bibfnamefont {J.~L.}\
  \bibnamefont {O'Brien}},\ }\href@noop {} {\bibfield  {journal} {\bibinfo
  {journal} {Science}\ }\textbf {\bibinfo {volume} {329}},\ \bibinfo {pages}
  {1500} (\bibinfo {year} {2010})}\BibitemShut {NoStop}%
\bibitem [{\citenamefont {Shadbolt}\ \emph {et~al.}(2011)\citenamefont
  {Shadbolt}, \citenamefont {Verde}, \citenamefont {Peruzzo}, \citenamefont
  {Politi}, \citenamefont {Laing}, \citenamefont {Lobino}, \citenamefont
  {Matthews}, \citenamefont {Thompson},\ and\ \citenamefont
  {O'Brien}}]{ShadboltNatP11}%
  \BibitemOpen
  \bibfield  {author} {\bibinfo {author} {\bibfnamefont {P.~J.}\ \bibnamefont
  {Shadbolt}}, \bibinfo {author} {\bibfnamefont {M.~R.}\ \bibnamefont {Verde}},
  \bibinfo {author} {\bibfnamefont {A.}~\bibnamefont {Peruzzo}}, \bibinfo
  {author} {\bibfnamefont {A.}~\bibnamefont {Politi}}, \bibinfo {author}
  {\bibfnamefont {A.}~\bibnamefont {Laing}}, \bibinfo {author} {\bibfnamefont
  {M.}~\bibnamefont {Lobino}}, \bibinfo {author} {\bibfnamefont {J.~C.~F.}\
  \bibnamefont {Matthews}}, \bibinfo {author} {\bibfnamefont {M.~G.}\
  \bibnamefont {Thompson}}, \ and\ \bibinfo {author} {\bibfnamefont {J.~L.}\
  \bibnamefont {O'Brien}},\ }\href@noop {} {\bibfield  {journal} {\bibinfo
  {journal} {Nat. Photon.}\ }\textbf {\bibinfo {volume} {6}},\ \bibinfo {pages}
  {45} (\bibinfo {year} {2011})}\BibitemShut {NoStop}%
\bibitem [{\citenamefont {Silverstone}\ \emph {et~al.}(2014)\citenamefont
  {Silverstone}, \citenamefont {Bonneau}, \citenamefont {Ohira}, \citenamefont
  {Suzuki}, \citenamefont {Yoshida}, \citenamefont {Iizuka}, \citenamefont
  {Ezaki}, \citenamefont {Natarajan}, \citenamefont {Tanner}, \citenamefont
  {Hadfield} \emph {et~al.}}]{SilverstoneNPhoton14}%
  \BibitemOpen
  \bibfield  {author} {\bibinfo {author} {\bibfnamefont {J.~W.}\ \bibnamefont
  {Silverstone}}, \bibinfo {author} {\bibfnamefont {D.}~\bibnamefont
  {Bonneau}}, \bibinfo {author} {\bibfnamefont {K.}~\bibnamefont {Ohira}},
  \bibinfo {author} {\bibfnamefont {N.}~\bibnamefont {Suzuki}}, \bibinfo
  {author} {\bibfnamefont {H.}~\bibnamefont {Yoshida}}, \bibinfo {author}
  {\bibfnamefont {N.}~\bibnamefont {Iizuka}}, \bibinfo {author} {\bibfnamefont
  {M.}~\bibnamefont {Ezaki}}, \bibinfo {author} {\bibfnamefont {C.~M.}\
  \bibnamefont {Natarajan}}, \bibinfo {author} {\bibfnamefont {M.~G.}\
  \bibnamefont {Tanner}}, \bibinfo {author} {\bibfnamefont {R.~H.}\
  \bibnamefont {Hadfield}},  \emph {et~al.},\ }\href@noop {} {\bibfield
  {journal} {\bibinfo  {journal} {Nat. Photon.}\ }\textbf {\bibinfo {volume}
  {8}},\ \bibinfo {pages} {104} (\bibinfo {year} {2014})}\BibitemShut {NoStop}%
\bibitem [{\citenamefont {O'Brien}\ \emph {et~al.}(2009)\citenamefont
  {O'Brien}, \citenamefont {Furusawa},\ and\ \citenamefont
  {Vu{\v{c}}kovi{\'c}}}]{OBrienNP09}%
  \BibitemOpen
  \bibfield  {author} {\bibinfo {author} {\bibfnamefont {J.~L.}\ \bibnamefont
  {O'Brien}}, \bibinfo {author} {\bibfnamefont {A.}~\bibnamefont {Furusawa}}, \
  and\ \bibinfo {author} {\bibfnamefont {J.}~\bibnamefont
  {Vu{\v{c}}kovi{\'c}}},\ }\href@noop {} {\bibfield  {journal} {\bibinfo
  {journal} {Nat. Photon.}\ }\textbf {\bibinfo {volume} {3}},\ \bibinfo {pages}
  {687} (\bibinfo {year} {2009})}\BibitemShut {NoStop}%
\bibitem [{\citenamefont {Politi}\ \emph {et~al.}(2008)\citenamefont {Politi},
  \citenamefont {Cryan}, \citenamefont {Rarity}, \citenamefont {Yu},\ and\
  \citenamefont {O'Brien}}]{PolitiScience08}%
  \BibitemOpen
  \bibfield  {author} {\bibinfo {author} {\bibfnamefont {A.}~\bibnamefont
  {Politi}}, \bibinfo {author} {\bibfnamefont {M.~J.}\ \bibnamefont {Cryan}},
  \bibinfo {author} {\bibfnamefont {J.~G.}\ \bibnamefont {Rarity}}, \bibinfo
  {author} {\bibfnamefont {S.}~\bibnamefont {Yu}}, \ and\ \bibinfo {author}
  {\bibfnamefont {J.~L.}\ \bibnamefont {O'Brien}},\ }\href@noop {} {\bibfield
  {journal} {\bibinfo  {journal} {Science}\ }\textbf {\bibinfo {volume}
  {320}},\ \bibinfo {pages} {646} (\bibinfo {year} {2008})}\BibitemShut
  {NoStop}%
\bibitem [{\citenamefont {Okamoto}\ \emph {et~al.}(2011)\citenamefont
  {Okamoto}, \citenamefont {O'Brien}, \citenamefont {Hofmann},\ and\
  \citenamefont {Takeuchi}}]{OkamotoPNAS11}%
  \BibitemOpen
  \bibfield  {author} {\bibinfo {author} {\bibfnamefont {R.}~\bibnamefont
  {Okamoto}}, \bibinfo {author} {\bibfnamefont {J.~L.}\ \bibnamefont
  {O'Brien}}, \bibinfo {author} {\bibfnamefont {H.~F.}\ \bibnamefont
  {Hofmann}}, \ and\ \bibinfo {author} {\bibfnamefont {S.}~\bibnamefont
  {Takeuchi}},\ }\href@noop {} {\bibfield  {journal} {\bibinfo  {journal}
  {Proc. Nat. Acad. Sci. (USA)}\ }\textbf {\bibinfo {volume} {108}},\ \bibinfo
  {pages} {10067} (\bibinfo {year} {2011})}\BibitemShut {NoStop}%
\bibitem [{\citenamefont {Crespi}\ \emph {et~al.}(2011)\citenamefont {Crespi},
  \citenamefont {Ramponi}, \citenamefont {Osellame}, \citenamefont {Sansoni},
  \citenamefont {Bongioanni}, \citenamefont {Sciarrino}, \citenamefont
  {Vallone},\ and\ \citenamefont {Mataloni}}]{CrespiNComm11}%
  \BibitemOpen
  \bibfield  {author} {\bibinfo {author} {\bibfnamefont {A.}~\bibnamefont
  {Crespi}}, \bibinfo {author} {\bibfnamefont {R.}~\bibnamefont {Ramponi}},
  \bibinfo {author} {\bibfnamefont {R.}~\bibnamefont {Osellame}}, \bibinfo
  {author} {\bibfnamefont {L.}~\bibnamefont {Sansoni}}, \bibinfo {author}
  {\bibfnamefont {I.}~\bibnamefont {Bongioanni}}, \bibinfo {author}
  {\bibfnamefont {F.}~\bibnamefont {Sciarrino}}, \bibinfo {author}
  {\bibfnamefont {G.}~\bibnamefont {Vallone}}, \ and\ \bibinfo {author}
  {\bibfnamefont {P.}~\bibnamefont {Mataloni}},\ }\href@noop {} {\bibfield
  {journal} {\bibinfo  {journal} {Nat. Commun.}\ }\textbf {\bibinfo {volume}
  {2}},\ \bibinfo {pages} {566} (\bibinfo {year} {2011})}\BibitemShut {NoStop}%
\bibitem [{\citenamefont {Bonneau}\ \emph {et~al.}(2012)\citenamefont
  {Bonneau}, \citenamefont {Engin}, \citenamefont {Ohira}, \citenamefont
  {Suzuki}, \citenamefont {Yoshida}, \citenamefont {Iizuka}, \citenamefont
  {Ezaki}, \citenamefont {Natarajan}, \citenamefont {Tanner}, \citenamefont
  {Hadfield}, \citenamefont {Dorenbos}, \citenamefont {Zwiller}, \citenamefont
  {O'Brien},\ and\ \citenamefont {Thompson}}]{BonneauNJPhys12}%
  \BibitemOpen
  \bibfield  {author} {\bibinfo {author} {\bibfnamefont {D.}~\bibnamefont
  {Bonneau}}, \bibinfo {author} {\bibfnamefont {E.}~\bibnamefont {Engin}},
  \bibinfo {author} {\bibfnamefont {K.}~\bibnamefont {Ohira}}, \bibinfo
  {author} {\bibfnamefont {N.}~\bibnamefont {Suzuki}}, \bibinfo {author}
  {\bibfnamefont {H.}~\bibnamefont {Yoshida}}, \bibinfo {author} {\bibfnamefont
  {N.}~\bibnamefont {Iizuka}}, \bibinfo {author} {\bibfnamefont
  {M.}~\bibnamefont {Ezaki}}, \bibinfo {author} {\bibfnamefont {C.~M.}\
  \bibnamefont {Natarajan}}, \bibinfo {author} {\bibfnamefont {M.~G.}\
  \bibnamefont {Tanner}}, \bibinfo {author} {\bibfnamefont {R.~H.}\
  \bibnamefont {Hadfield}}, \bibinfo {author} {\bibfnamefont {S.~N.}\
  \bibnamefont {Dorenbos}}, \bibinfo {author} {\bibfnamefont {V.}~\bibnamefont
  {Zwiller}}, \bibinfo {author} {\bibfnamefont {J.~L.}\ \bibnamefont
  {O'Brien}}, \ and\ \bibinfo {author} {\bibfnamefont {M.~G.}\ \bibnamefont
  {Thompson}},\ }\href@noop {} {\bibfield  {journal} {\bibinfo  {journal} {New
  J. Phys.}\ }\textbf {\bibinfo {volume} {14}},\ \bibinfo {pages} {045003}
  (\bibinfo {year} {2012})}\BibitemShut {NoStop}%
\bibitem [{\citenamefont {Li}\ \emph {et~al.}(2011)\citenamefont {Li},
  \citenamefont {Przeslak}, \citenamefont {Niskanen}, \citenamefont {Matthews},
  \citenamefont {Politi}, \citenamefont {Shadbolt}, \citenamefont {Laing},
  \citenamefont {Lobino}, \citenamefont {Thompson},\ and\ \citenamefont
  {O'Brien}}]{HWLiNJPhys11}%
  \BibitemOpen
  \bibfield  {author} {\bibinfo {author} {\bibfnamefont {H.~W.}\ \bibnamefont
  {Li}}, \bibinfo {author} {\bibfnamefont {S.}~\bibnamefont {Przeslak}},
  \bibinfo {author} {\bibfnamefont {A.~O.}\ \bibnamefont {Niskanen}}, \bibinfo
  {author} {\bibfnamefont {J.~C.~F.}\ \bibnamefont {Matthews}}, \bibinfo
  {author} {\bibfnamefont {A.}~\bibnamefont {Politi}}, \bibinfo {author}
  {\bibfnamefont {P.}~\bibnamefont {Shadbolt}}, \bibinfo {author}
  {\bibfnamefont {A.}~\bibnamefont {Laing}}, \bibinfo {author} {\bibfnamefont
  {M.}~\bibnamefont {Lobino}}, \bibinfo {author} {\bibfnamefont {M.~G.}\
  \bibnamefont {Thompson}}, \ and\ \bibinfo {author} {\bibfnamefont {J.~L.}\
  \bibnamefont {O'Brien}},\ }\href@noop {} {\bibfield  {journal} {\bibinfo
  {journal} {New J. Phys.}\ }\textbf {\bibinfo {volume} {13}},\ \bibinfo
  {pages} {115009} (\bibinfo {year} {2011})}\BibitemShut {NoStop}%
\bibitem [{\citenamefont {Metcalf}\ \emph {et~al.}(2014)\citenamefont
  {Metcalf}, \citenamefont {Spring}, \citenamefont {Humphreys}, \citenamefont
  {Thomas-Peter}, \citenamefont {Barbieri}, \citenamefont {Kolthammer},
  \citenamefont {Jin}, \citenamefont {Langford}, \citenamefont {Kundys},
  \citenamefont {Gates} \emph {et~al.}}]{MetcalfNP14}%
  \BibitemOpen
  \bibfield  {author} {\bibinfo {author} {\bibfnamefont {B.~J.}\ \bibnamefont
  {Metcalf}}, \bibinfo {author} {\bibfnamefont {J.~B.}\ \bibnamefont {Spring}},
  \bibinfo {author} {\bibfnamefont {P.~C.}\ \bibnamefont {Humphreys}}, \bibinfo
  {author} {\bibfnamefont {N.}~\bibnamefont {Thomas-Peter}}, \bibinfo {author}
  {\bibfnamefont {M.}~\bibnamefont {Barbieri}}, \bibinfo {author}
  {\bibfnamefont {W.~S.}\ \bibnamefont {Kolthammer}}, \bibinfo {author}
  {\bibfnamefont {X.-M.}\ \bibnamefont {Jin}}, \bibinfo {author} {\bibfnamefont
  {N.~K.}\ \bibnamefont {Langford}}, \bibinfo {author} {\bibfnamefont
  {D.}~\bibnamefont {Kundys}}, \bibinfo {author} {\bibfnamefont {J.~C.}\
  \bibnamefont {Gates}},  \emph {et~al.},\ }\href@noop {} {\bibfield  {journal}
  {\bibinfo  {journal} {Nat. Photon.}\ }\textbf {\bibinfo {volume} {8}},\
  \bibinfo {pages} {770} (\bibinfo {year} {2014})}\BibitemShut {NoStop}%
\bibitem [{\citenamefont {Perets}\ \emph {et~al.}(2008)\citenamefont {Perets},
  \citenamefont {Lahini}, \citenamefont {Pozzi}, \citenamefont {Sorel},
  \citenamefont {Morandotti},\ and\ \citenamefont {Silberberg}}]{PeretsPRL08}%
  \BibitemOpen
  \bibfield  {author} {\bibinfo {author} {\bibfnamefont {H.~B.}\ \bibnamefont
  {Perets}}, \bibinfo {author} {\bibfnamefont {Y.}~\bibnamefont {Lahini}},
  \bibinfo {author} {\bibfnamefont {F.}~\bibnamefont {Pozzi}}, \bibinfo
  {author} {\bibfnamefont {M.}~\bibnamefont {Sorel}}, \bibinfo {author}
  {\bibfnamefont {R.}~\bibnamefont {Morandotti}}, \ and\ \bibinfo {author}
  {\bibfnamefont {Y.}~\bibnamefont {Silberberg}},\ }\href@noop {} {\bibfield
  {journal} {\bibinfo  {journal} {Phys. Rev. Lett.}\ }\textbf {\bibinfo
  {volume} {100}},\ \bibinfo {pages} {170506} (\bibinfo {year}
  {2008})}\BibitemShut {NoStop}%
\bibitem [{\citenamefont {Rai}\ \emph {et~al.}(2008)\citenamefont {Rai},
  \citenamefont {Agarwal},\ and\ \citenamefont {Perk}}]{RaiPRA08}%
  \BibitemOpen
  \bibfield  {author} {\bibinfo {author} {\bibfnamefont {A.}~\bibnamefont
  {Rai}}, \bibinfo {author} {\bibfnamefont {G.~S.}\ \bibnamefont {Agarwal}}, \
  and\ \bibinfo {author} {\bibfnamefont {J.~H.~H.}\ \bibnamefont {Perk}},\
  }\href@noop {} {\bibfield  {journal} {\bibinfo  {journal} {Phys. Rev. A}\
  }\textbf {\bibinfo {volume} {78}},\ \bibinfo {pages} {042304} (\bibinfo
  {year} {2008})}\BibitemShut {NoStop}%
\bibitem [{\citenamefont {Abouraddy}\ \emph {et~al.}(2012)\citenamefont
  {Abouraddy}, \citenamefont {{Di G}iuseppe}, \citenamefont {Christodoulides},\
  and\ \citenamefont {Saleh}}]{AbouraddyPRA12}%
  \BibitemOpen
  \bibfield  {author} {\bibinfo {author} {\bibfnamefont {A.~F.}\ \bibnamefont
  {Abouraddy}}, \bibinfo {author} {\bibfnamefont {G.}~\bibnamefont {{Di
  G}iuseppe}}, \bibinfo {author} {\bibfnamefont {D.~N.}\ \bibnamefont
  {Christodoulides}}, \ and\ \bibinfo {author} {\bibfnamefont {B.~E.~A.}\
  \bibnamefont {Saleh}},\ }\href@noop {} {\bibfield  {journal} {\bibinfo
  {journal} {Phys. Rev. A}\ }\textbf {\bibinfo {volume} {86}},\ \bibinfo
  {pages} {040302 (R)} (\bibinfo {year} {2012})}\BibitemShut {NoStop}%
\bibitem [{\citenamefont {{Di G}iuseppe}\ \emph {et~al.}(2013)\citenamefont
  {{Di G}iuseppe}, \citenamefont {Martin}, \citenamefont {Perez-Leija},
  \citenamefont {Keil}, \citenamefont {Dreisow}, \citenamefont {Nolte},
  \citenamefont {Szameit}, \citenamefont {Abouraddy}, \citenamefont
  {Christodoulides},\ and\ \citenamefont {Saleh}}]{DiGiuseppePRL13}%
  \BibitemOpen
  \bibfield  {author} {\bibinfo {author} {\bibfnamefont {G.}~\bibnamefont {{Di
  G}iuseppe}}, \bibinfo {author} {\bibfnamefont {L.}~\bibnamefont {Martin}},
  \bibinfo {author} {\bibfnamefont {A.}~\bibnamefont {Perez-Leija}}, \bibinfo
  {author} {\bibfnamefont {R.}~\bibnamefont {Keil}}, \bibinfo {author}
  {\bibfnamefont {F.}~\bibnamefont {Dreisow}}, \bibinfo {author} {\bibfnamefont
  {S.}~\bibnamefont {Nolte}}, \bibinfo {author} {\bibfnamefont
  {A.}~\bibnamefont {Szameit}}, \bibinfo {author} {\bibfnamefont {A.~F.}\
  \bibnamefont {Abouraddy}}, \bibinfo {author} {\bibfnamefont {D.~N.}\
  \bibnamefont {Christodoulides}}, \ and\ \bibinfo {author} {\bibfnamefont
  {B.~E.~A.}\ \bibnamefont {Saleh}},\ }\href@noop {} {\bibfield  {journal}
  {\bibinfo  {journal} {Phys. Rev. Lett.}\ }\textbf {\bibinfo {volume} {110}},\
  \bibinfo {pages} {150503} (\bibinfo {year} {2013})}\BibitemShut {NoStop}%
\bibitem [{\citenamefont {Gilead}\ \emph {et~al.}(2015)\citenamefont {Gilead},
  \citenamefont {Verbin},\ and\ \citenamefont {Silberberg}}]{GileadPRL15}%
  \BibitemOpen
  \bibfield  {author} {\bibinfo {author} {\bibfnamefont {Y.}~\bibnamefont
  {Gilead}}, \bibinfo {author} {\bibfnamefont {M.}~\bibnamefont {Verbin}}, \
  and\ \bibinfo {author} {\bibfnamefont {Y.}~\bibnamefont {Silberberg}},\
  }\href@noop {} {\bibfield  {journal} {\bibinfo  {journal} {Phys. Rev. Lett.}\
  }\textbf {\bibinfo {volume} {115}},\ \bibinfo {pages} {133602} (\bibinfo
  {year} {2015})}\BibitemShut {NoStop}%
\bibitem [{\citenamefont {Horn}\ \emph {et~al.}(2012)\citenamefont {Horn},
  \citenamefont {Abolghasem}, \citenamefont {Bijlani}, \citenamefont {Kang},
  \citenamefont {Helmy},\ and\ \citenamefont {Weihs}}]{HornPRL12}%
  \BibitemOpen
  \bibfield  {author} {\bibinfo {author} {\bibfnamefont {R.}~\bibnamefont
  {Horn}}, \bibinfo {author} {\bibfnamefont {P.}~\bibnamefont {Abolghasem}},
  \bibinfo {author} {\bibfnamefont {B.~J.}\ \bibnamefont {Bijlani}}, \bibinfo
  {author} {\bibfnamefont {D.}~\bibnamefont {Kang}}, \bibinfo {author}
  {\bibfnamefont {A.}~\bibnamefont {Helmy}}, \ and\ \bibinfo {author}
  {\bibfnamefont {G.}~\bibnamefont {Weihs}},\ }\href@noop {} {\bibfield
  {journal} {\bibinfo  {journal} {Phys. Rev. Lett.}\ }\textbf {\bibinfo
  {volume} {108}},\ \bibinfo {pages} {153605} (\bibinfo {year}
  {2012})}\BibitemShut {NoStop}%
\bibitem [{\citenamefont {Orieux}\ \emph {et~al.}(2013)\citenamefont {Orieux},
  \citenamefont {Eckstein}, \citenamefont {Lema{\^\i}tre}, \citenamefont
  {Filloux}, \citenamefont {Favero}, \citenamefont {Leo}, \citenamefont
  {Coudreau}, \citenamefont {Keller}, \citenamefont {Milman},\ and\
  \citenamefont {Ducci}}]{OrieuxPRL13}%
  \BibitemOpen
  \bibfield  {author} {\bibinfo {author} {\bibfnamefont {A.}~\bibnamefont
  {Orieux}}, \bibinfo {author} {\bibfnamefont {A.}~\bibnamefont {Eckstein}},
  \bibinfo {author} {\bibfnamefont {A.}~\bibnamefont {Lema{\^\i}tre}}, \bibinfo
  {author} {\bibfnamefont {P.}~\bibnamefont {Filloux}}, \bibinfo {author}
  {\bibfnamefont {I.}~\bibnamefont {Favero}}, \bibinfo {author} {\bibfnamefont
  {G.}~\bibnamefont {Leo}}, \bibinfo {author} {\bibfnamefont {T.}~\bibnamefont
  {Coudreau}}, \bibinfo {author} {\bibfnamefont {A.}~\bibnamefont {Keller}},
  \bibinfo {author} {\bibfnamefont {P.}~\bibnamefont {Milman}}, \ and\ \bibinfo
  {author} {\bibfnamefont {S.}~\bibnamefont {Ducci}},\ }\href@noop {}
  {\bibfield  {journal} {\bibinfo  {journal} {Phys. Rev. Lett.}\ }\textbf
  {\bibinfo {volume} {110}},\ \bibinfo {pages} {160502} (\bibinfo {year}
  {2013})}\BibitemShut {NoStop}%
\bibitem [{\citenamefont {Boitier}\ \emph {et~al.}(2014)\citenamefont
  {Boitier}, \citenamefont {Orieux}, \citenamefont {Autebert}, \citenamefont
  {Lema{\^\i}tre}, \citenamefont {Galopin}, \citenamefont {Manquest},
  \citenamefont {Sirtori}, \citenamefont {Favero}, \citenamefont {Leo},\ and\
  \citenamefont {Ducci}}]{BoitierPRL14}%
  \BibitemOpen
  \bibfield  {author} {\bibinfo {author} {\bibfnamefont {F.}~\bibnamefont
  {Boitier}}, \bibinfo {author} {\bibfnamefont {A.}~\bibnamefont {Orieux}},
  \bibinfo {author} {\bibfnamefont {C.}~\bibnamefont {Autebert}}, \bibinfo
  {author} {\bibfnamefont {A.}~\bibnamefont {Lema{\^\i}tre}}, \bibinfo {author}
  {\bibfnamefont {E.}~\bibnamefont {Galopin}}, \bibinfo {author} {\bibfnamefont
  {C.}~\bibnamefont {Manquest}}, \bibinfo {author} {\bibfnamefont
  {C.}~\bibnamefont {Sirtori}}, \bibinfo {author} {\bibfnamefont
  {I.}~\bibnamefont {Favero}}, \bibinfo {author} {\bibfnamefont
  {G.}~\bibnamefont {Leo}}, \ and\ \bibinfo {author} {\bibfnamefont
  {S.}~\bibnamefont {Ducci}},\ }\href@noop {} {\bibfield  {journal} {\bibinfo
  {journal} {Phys. Rev. Lett.}\ }\textbf {\bibinfo {volume} {112}},\ \bibinfo
  {pages} {183901} (\bibinfo {year} {2014})}\BibitemShut {NoStop}%
\bibitem [{\citenamefont {Matsuda}\ \emph {et~al.}(2012)\citenamefont
  {Matsuda}, \citenamefont {Le~Jeannic}, \citenamefont {Fukuda}, \citenamefont
  {Tsuchizawa}, \citenamefont {Munro}, \citenamefont {Shimizu}, \citenamefont
  {Yamada}, \citenamefont {Tokura},\ and\ \citenamefont
  {Takesue}}]{MatsudaSCiRep12}%
  \BibitemOpen
  \bibfield  {author} {\bibinfo {author} {\bibfnamefont {N.}~\bibnamefont
  {Matsuda}}, \bibinfo {author} {\bibfnamefont {H.}~\bibnamefont {Le~Jeannic}},
  \bibinfo {author} {\bibfnamefont {H.}~\bibnamefont {Fukuda}}, \bibinfo
  {author} {\bibfnamefont {T.}~\bibnamefont {Tsuchizawa}}, \bibinfo {author}
  {\bibfnamefont {W.~J.}\ \bibnamefont {Munro}}, \bibinfo {author}
  {\bibfnamefont {K.}~\bibnamefont {Shimizu}}, \bibinfo {author} {\bibfnamefont
  {K.}~\bibnamefont {Yamada}}, \bibinfo {author} {\bibfnamefont
  {Y.}~\bibnamefont {Tokura}}, \ and\ \bibinfo {author} {\bibfnamefont
  {H.}~\bibnamefont {Takesue}},\ }\href@noop {} {\bibfield  {journal} {\bibinfo
   {journal} {Sci. Rep.}\ }\textbf {\bibinfo {volume} {2}} (\bibinfo {year}
  {2012})}\BibitemShut {NoStop}%
\bibitem [{\citenamefont {Christodoulides}\ \emph {et~al.}(2003)\citenamefont
  {Christodoulides}, \citenamefont {Lederer},\ and\ \citenamefont
  {Silberberg}}]{ChristodoulidesNat03}%
  \BibitemOpen
  \bibfield  {author} {\bibinfo {author} {\bibfnamefont {D.~N.}\ \bibnamefont
  {Christodoulides}}, \bibinfo {author} {\bibfnamefont {F.}~\bibnamefont
  {Lederer}}, \ and\ \bibinfo {author} {\bibfnamefont {Y.}~\bibnamefont
  {Silberberg}},\ }\href@noop {} {\bibfield  {journal} {\bibinfo  {journal}
  {Nature}\ }\textbf {\bibinfo {volume} {424}},\ \bibinfo {pages} {817}
  (\bibinfo {year} {2003})}\BibitemShut {NoStop}%
\bibitem [{\citenamefont {Lederer}\ \emph {et~al.}(2008)\citenamefont
  {Lederer}, \citenamefont {Stegeman}, \citenamefont {Christodoulides},
  \citenamefont {Assanto}, \citenamefont {Segev},\ and\ \citenamefont
  {Silberberg}}]{LedererPhysRep08}%
  \BibitemOpen
  \bibfield  {author} {\bibinfo {author} {\bibfnamefont {F.}~\bibnamefont
  {Lederer}}, \bibinfo {author} {\bibfnamefont {G.~I.}\ \bibnamefont
  {Stegeman}}, \bibinfo {author} {\bibfnamefont {D.~N.}\ \bibnamefont
  {Christodoulides}}, \bibinfo {author} {\bibfnamefont {G.}~\bibnamefont
  {Assanto}}, \bibinfo {author} {\bibfnamefont {M.}~\bibnamefont {Segev}}, \
  and\ \bibinfo {author} {\bibfnamefont {Y.}~\bibnamefont {Silberberg}},\
  }\href@noop {} {\bibfield  {journal} {\bibinfo  {journal} {Phys. Rep.}\
  }\textbf {\bibinfo {volume} {463}},\ \bibinfo {pages} {1} (\bibinfo {year}
  {2008})}\BibitemShut {NoStop}%
\bibitem [{\citenamefont {Kartashov}\ \emph {et~al.}(2011)\citenamefont
  {Kartashov}, \citenamefont {Malomed},\ and\ \citenamefont
  {Torner}}]{KartashovRMP11}%
  \BibitemOpen
  \bibfield  {author} {\bibinfo {author} {\bibfnamefont {Y.~V.}\ \bibnamefont
  {Kartashov}}, \bibinfo {author} {\bibfnamefont {B.~A.}\ \bibnamefont
  {Malomed}}, \ and\ \bibinfo {author} {\bibfnamefont {L.}~\bibnamefont
  {Torner}},\ }\href@noop {} {\bibfield  {journal} {\bibinfo  {journal} {Rev.
  Mod. Phys.}\ }\textbf {\bibinfo {volume} {83}},\ \bibinfo {pages} {247}
  (\bibinfo {year} {2011})}\BibitemShut {NoStop}%
\bibitem [{\citenamefont {Flach}\ and\ \citenamefont
  {Gorbach}(2008)}]{FlachPhysRep08}%
  \BibitemOpen
  \bibfield  {author} {\bibinfo {author} {\bibfnamefont {S.}~\bibnamefont
  {Flach}}\ and\ \bibinfo {author} {\bibfnamefont {A.~V.}\ \bibnamefont
  {Gorbach}},\ }\href@noop {} {\bibfield  {journal} {\bibinfo  {journal} {Phys.
  Rep.}\ }\textbf {\bibinfo {volume} {467}},\ \bibinfo {pages} {1} (\bibinfo
  {year} {2008})}\BibitemShut {NoStop}%
\bibitem [{\citenamefont {Malomed}\ \emph {et~al.}(2005)\citenamefont
  {Malomed}, \citenamefont {Mihalache}, \citenamefont {Wise},\ and\
  \citenamefont {Torner}}]{MalomedJOptB05}%
  \BibitemOpen
  \bibfield  {author} {\bibinfo {author} {\bibfnamefont {B.~A.}\ \bibnamefont
  {Malomed}}, \bibinfo {author} {\bibfnamefont {D.}~\bibnamefont {Mihalache}},
  \bibinfo {author} {\bibfnamefont {F.}~\bibnamefont {Wise}}, \ and\ \bibinfo
  {author} {\bibfnamefont {L.}~\bibnamefont {Torner}},\ }\href@noop {}
  {\bibfield  {journal} {\bibinfo  {journal} {Journal of Optics B: Quantum and
  Semiclassical Optics}\ }\textbf {\bibinfo {volume} {7}},\ \bibinfo {pages}
  {R53} (\bibinfo {year} {2005})}\BibitemShut {NoStop}%
\bibitem [{\citenamefont {Solntsev}\ \emph {et~al.}(2012)\citenamefont
  {Solntsev}, \citenamefont {Sukhorukov}, \citenamefont {Neshev},\ and\
  \citenamefont {Kivshar}}]{SolntsevPRL12}%
  \BibitemOpen
  \bibfield  {author} {\bibinfo {author} {\bibfnamefont {A.~S.}\ \bibnamefont
  {Solntsev}}, \bibinfo {author} {\bibfnamefont {A.~A.}\ \bibnamefont
  {Sukhorukov}}, \bibinfo {author} {\bibfnamefont {D.~N.}\ \bibnamefont
  {Neshev}}, \ and\ \bibinfo {author} {\bibfnamefont {Y.~S.}\ \bibnamefont
  {Kivshar}},\ }\href@noop {} {\bibfield  {journal} {\bibinfo  {journal} {Phys.
  Rev. Lett.}\ }\textbf {\bibinfo {volume} {108}},\ \bibinfo {pages} {023601}
  (\bibinfo {year} {2012})}\BibitemShut {NoStop}%
\bibitem [{\citenamefont {Rarity}\ and\ \citenamefont
  {Tapster}(1990{\natexlab{a}})}]{RarityPRL90}%
  \BibitemOpen
  \bibfield  {author} {\bibinfo {author} {\bibfnamefont {J.}~\bibnamefont
  {Rarity}}\ and\ \bibinfo {author} {\bibfnamefont {P.}~\bibnamefont
  {Tapster}},\ }\href@noop {} {\bibfield  {journal} {\bibinfo  {journal} {Phys.
  Rev. Lett.}\ }\textbf {\bibinfo {volume} {64}},\ \bibinfo {pages} {2495}
  (\bibinfo {year} {1990}{\natexlab{a}})}\BibitemShut {NoStop}%
\bibitem [{\citenamefont {Rarity}\ and\ \citenamefont
  {Tapster}(1990{\natexlab{b}})}]{RarityPRA90}%
  \BibitemOpen
  \bibfield  {author} {\bibinfo {author} {\bibfnamefont {J.}~\bibnamefont
  {Rarity}}\ and\ \bibinfo {author} {\bibfnamefont {P.}~\bibnamefont
  {Tapster}},\ }\href@noop {} {\bibfield  {journal} {\bibinfo  {journal} {Phys.
  Rev. A}\ }\textbf {\bibinfo {volume} {41}},\ \bibinfo {pages} {5139}
  (\bibinfo {year} {1990}{\natexlab{b}})}\BibitemShut {NoStop}%
\bibitem [{\citenamefont {Saleh}\ \emph {et~al.}(2000)\citenamefont {Saleh},
  \citenamefont {Abouraddy}, \citenamefont {Sergienko},\ and\ \citenamefont
  {Teich}}]{SalehPRA00}%
  \BibitemOpen
  \bibfield  {author} {\bibinfo {author} {\bibfnamefont {B.~E.}\ \bibnamefont
  {Saleh}}, \bibinfo {author} {\bibfnamefont {A.~F.}\ \bibnamefont
  {Abouraddy}}, \bibinfo {author} {\bibfnamefont {A.~V.}\ \bibnamefont
  {Sergienko}}, \ and\ \bibinfo {author} {\bibfnamefont {M.~C.}\ \bibnamefont
  {Teich}},\ }\href@noop {} {\bibfield  {journal} {\bibinfo  {journal} {Phys.
  Rev. A}\ }\textbf {\bibinfo {volume} {62}},\ \bibinfo {pages} {043816}
  (\bibinfo {year} {2000})}\BibitemShut {NoStop}%
\bibitem [{\citenamefont {Pan}\ \emph {et~al.}(2001)\citenamefont {Pan},
  \citenamefont {Daniell}, \citenamefont {Gasparoni}, \citenamefont {Weihs},\
  and\ \citenamefont {Zeilinger}}]{PanPRL01}%
  \BibitemOpen
  \bibfield  {author} {\bibinfo {author} {\bibfnamefont {J.-W.}\ \bibnamefont
  {Pan}}, \bibinfo {author} {\bibfnamefont {M.}~\bibnamefont {Daniell}},
  \bibinfo {author} {\bibfnamefont {S.}~\bibnamefont {Gasparoni}}, \bibinfo
  {author} {\bibfnamefont {G.}~\bibnamefont {Weihs}}, \ and\ \bibinfo {author}
  {\bibfnamefont {A.}~\bibnamefont {Zeilinger}},\ }\href@noop {} {\bibfield
  {journal} {\bibinfo  {journal} {Phys. Rev. Lett.}\ }\textbf {\bibinfo
  {volume} {86}},\ \bibinfo {pages} {4435} (\bibinfo {year}
  {2001})}\BibitemShut {NoStop}%
\bibitem [{\citenamefont {Chen}\ \emph {et~al.}(2003)\citenamefont {Chen},
  \citenamefont {Pan}, \citenamefont {Zhang}, \citenamefont {Brukner},\ and\
  \citenamefont {Zeilinger}}]{ChenPRL03}%
  \BibitemOpen
  \bibfield  {author} {\bibinfo {author} {\bibfnamefont {Z.-B.}\ \bibnamefont
  {Chen}}, \bibinfo {author} {\bibfnamefont {J.-W.}\ \bibnamefont {Pan}},
  \bibinfo {author} {\bibfnamefont {Y.-D.}\ \bibnamefont {Zhang}}, \bibinfo
  {author} {\bibfnamefont {{\v{C}}.}~\bibnamefont {Brukner}}, \ and\ \bibinfo
  {author} {\bibfnamefont {A.}~\bibnamefont {Zeilinger}},\ }\href@noop {}
  {\bibfield  {journal} {\bibinfo  {journal} {Phys. Rev. Lett.}\ }\textbf
  {\bibinfo {volume} {90}},\ \bibinfo {pages} {160408} (\bibinfo {year}
  {2003})}\BibitemShut {NoStop}%
\bibitem [{\citenamefont {Hong}\ \emph {et~al.}(1987)\citenamefont {Hong},
  \citenamefont {Ou},\ and\ \citenamefont {Mandel}}]{HongPRL87}%
  \BibitemOpen
  \bibfield  {author} {\bibinfo {author} {\bibfnamefont {C.}~\bibnamefont
  {Hong}}, \bibinfo {author} {\bibfnamefont {Z.}~\bibnamefont {Ou}}, \ and\
  \bibinfo {author} {\bibfnamefont {L.}~\bibnamefont {Mandel}},\ }\href@noop {}
  {\bibfield  {journal} {\bibinfo  {journal} {Phys. Rev. Lett.}\ }\textbf
  {\bibinfo {volume} {59}},\ \bibinfo {pages} {2044} (\bibinfo {year}
  {1987})}\BibitemShut {NoStop}%
\bibitem [{\citenamefont {Zou}\ \emph {et~al.}(1991)\citenamefont {Zou},
  \citenamefont {Wang},\ and\ \citenamefont {Mandel}}]{ZouPRL91}%
  \BibitemOpen
  \bibfield  {author} {\bibinfo {author} {\bibfnamefont {X.}~\bibnamefont
  {Zou}}, \bibinfo {author} {\bibfnamefont {L.}~\bibnamefont {Wang}}, \ and\
  \bibinfo {author} {\bibfnamefont {L.}~\bibnamefont {Mandel}},\ }\href@noop {}
  {\bibfield  {journal} {\bibinfo  {journal} {Phys. Rev. Lett.}\ }\textbf
  {\bibinfo {volume} {67}},\ \bibinfo {pages} {318} (\bibinfo {year}
  {1991})}\BibitemShut {NoStop}%
\bibitem [{\citenamefont {Lemos}\ \emph {et~al.}(2014)\citenamefont {Lemos},
  \citenamefont {Borish}, \citenamefont {Cole}, \citenamefont {Ramelow},
  \citenamefont {Lapkiewicz},\ and\ \citenamefont {Zeilinger}}]{LemosNat14}%
  \BibitemOpen
  \bibfield  {author} {\bibinfo {author} {\bibfnamefont {G.~B.}\ \bibnamefont
  {Lemos}}, \bibinfo {author} {\bibfnamefont {V.}~\bibnamefont {Borish}},
  \bibinfo {author} {\bibfnamefont {G.~D.}\ \bibnamefont {Cole}}, \bibinfo
  {author} {\bibfnamefont {S.}~\bibnamefont {Ramelow}}, \bibinfo {author}
  {\bibfnamefont {R.}~\bibnamefont {Lapkiewicz}}, \ and\ \bibinfo {author}
  {\bibfnamefont {A.}~\bibnamefont {Zeilinger}},\ }\href@noop {} {\bibfield
  {journal} {\bibinfo  {journal} {Nature}\ }\textbf {\bibinfo {volume} {512}},\
  \bibinfo {pages} {409} (\bibinfo {year} {2014})}\BibitemShut {NoStop}%
\bibitem [{\citenamefont {Booth}\ \emph {et~al.}(2002)\citenamefont {Booth},
  \citenamefont {Atat{\"u}re}, \citenamefont {{Di G}iuseppe}, \citenamefont
  {Saleh}, \citenamefont {Sergienko},\ and\ \citenamefont
  {Teich}}]{BoothPRA02}%
  \BibitemOpen
  \bibfield  {author} {\bibinfo {author} {\bibfnamefont {M.~C.}\ \bibnamefont
  {Booth}}, \bibinfo {author} {\bibfnamefont {M.}~\bibnamefont {Atat{\"u}re}},
  \bibinfo {author} {\bibfnamefont {G.}~\bibnamefont {{Di G}iuseppe}}, \bibinfo
  {author} {\bibfnamefont {B.~E.~A.}\ \bibnamefont {Saleh}}, \bibinfo {author}
  {\bibfnamefont {A.~V.}\ \bibnamefont {Sergienko}}, \ and\ \bibinfo {author}
  {\bibfnamefont {M.~C.}\ \bibnamefont {Teich}},\ }\href@noop {} {\bibfield
  {journal} {\bibinfo  {journal} {Phys. Rev. A}\ }\textbf {\bibinfo {volume}
  {66}},\ \bibinfo {pages} {023815} (\bibinfo {year} {2002})}\BibitemShut
  {NoStop}%
\bibitem [{\citenamefont {{De R}ossi}\ and\ \citenamefont
  {Berger}(2002)}]{DeRossiPRL02}%
  \BibitemOpen
  \bibfield  {author} {\bibinfo {author} {\bibfnamefont {A.}~\bibnamefont {{De
  R}ossi}}\ and\ \bibinfo {author} {\bibfnamefont {V.}~\bibnamefont {Berger}},\
  }\href@noop {} {\bibfield  {journal} {\bibinfo  {journal} {Phys. Rev. Lett.}\
  }\textbf {\bibinfo {volume} {88}},\ \bibinfo {pages} {043901} (\bibinfo
  {year} {2002})}\BibitemShut {NoStop}%
\bibitem [{\citenamefont {Bortz}\ and\ \citenamefont
  {Fejer}(1991)}]{BortzOL91}%
  \BibitemOpen
  \bibfield  {author} {\bibinfo {author} {\bibfnamefont {M.}~\bibnamefont
  {Bortz}}\ and\ \bibinfo {author} {\bibfnamefont {M.}~\bibnamefont {Fejer}},\
  }\href@noop {} {\bibfield  {journal} {\bibinfo  {journal} {Opt. Lett.}\
  }\textbf {\bibinfo {volume} {16}},\ \bibinfo {pages} {1844} (\bibinfo {year}
  {1991})}\BibitemShut {NoStop}%
\bibitem [{\citenamefont {Abouraddy}\ \emph {et~al.}(2001)\citenamefont
  {Abouraddy}, \citenamefont {Saleh}, \citenamefont {Sergienko},\ and\
  \citenamefont {Teich}}]{AbouraddyPRA01}%
  \BibitemOpen
  \bibfield  {author} {\bibinfo {author} {\bibfnamefont {A.~F.}\ \bibnamefont
  {Abouraddy}}, \bibinfo {author} {\bibfnamefont {B.~E.}\ \bibnamefont
  {Saleh}}, \bibinfo {author} {\bibfnamefont {A.~V.}\ \bibnamefont
  {Sergienko}}, \ and\ \bibinfo {author} {\bibfnamefont {M.~C.}\ \bibnamefont
  {Teich}},\ }\href@noop {} {\bibfield  {journal} {\bibinfo  {journal} {Phys.
  Rev. A}\ }\textbf {\bibinfo {volume} {64}},\ \bibinfo {pages} {050101}
  (\bibinfo {year} {2001})}\BibitemShut {NoStop}%
\bibitem [{\citenamefont {Guarino}\ \emph {et~al.}(2007)\citenamefont
  {Guarino}, \citenamefont {Poberaj}, \citenamefont {Rezzonico}, \citenamefont
  {Degl'Innocenti},\ and\ \citenamefont {G{\"u}nter}}]{GuarinoNPhoton07}%
  \BibitemOpen
  \bibfield  {author} {\bibinfo {author} {\bibfnamefont {A.}~\bibnamefont
  {Guarino}}, \bibinfo {author} {\bibfnamefont {G.}~\bibnamefont {Poberaj}},
  \bibinfo {author} {\bibfnamefont {D.}~\bibnamefont {Rezzonico}}, \bibinfo
  {author} {\bibfnamefont {R.}~\bibnamefont {Degl'Innocenti}}, \ and\ \bibinfo
  {author} {\bibfnamefont {P.}~\bibnamefont {G{\"u}nter}},\ }\href@noop {}
  {\bibfield  {journal} {\bibinfo  {journal} {Nat. Photon.}\ }\textbf {\bibinfo
  {volume} {1}},\ \bibinfo {pages} {407} (\bibinfo {year} {2007})}\BibitemShut
  {NoStop}%
\bibitem [{\citenamefont {Rabelo}\ \emph {et~al.}(2011)\citenamefont {Rabelo},
  \citenamefont {Eknoyan},\ and\ \citenamefont {Taylor}}]{RabeloAO11}%
  \BibitemOpen
  \bibfield  {author} {\bibinfo {author} {\bibfnamefont {R.~C.}\ \bibnamefont
  {Rabelo}}, \bibinfo {author} {\bibfnamefont {O.}~\bibnamefont {Eknoyan}}, \
  and\ \bibinfo {author} {\bibfnamefont {H.~F.}\ \bibnamefont {Taylor}},\
  }\href@noop {} {\bibfield  {journal} {\bibinfo  {journal} {Appl. Opt.}\
  }\textbf {\bibinfo {volume} {50}},\ \bibinfo {pages} {562} (\bibinfo {year}
  {2011})}\BibitemShut {NoStop}%
\bibitem [{\citenamefont {Locatelli}\ \emph {et~al.}(2012)\citenamefont
  {Locatelli}, \citenamefont {Capobianco}, \citenamefont {Midrio},
  \citenamefont {Boscolo},\ and\ \citenamefont {De~Angelis}}]{LocatelliOE12}%
  \BibitemOpen
  \bibfield  {author} {\bibinfo {author} {\bibfnamefont {A.}~\bibnamefont
  {Locatelli}}, \bibinfo {author} {\bibfnamefont {A.-D.}\ \bibnamefont
  {Capobianco}}, \bibinfo {author} {\bibfnamefont {M.}~\bibnamefont {Midrio}},
  \bibinfo {author} {\bibfnamefont {S.}~\bibnamefont {Boscolo}}, \ and\
  \bibinfo {author} {\bibfnamefont {C.}~\bibnamefont {De~Angelis}},\
  }\href@noop {} {\bibfield  {journal} {\bibinfo  {journal} {Opt. Express}\
  }\textbf {\bibinfo {volume} {20}},\ \bibinfo {pages} {28479} (\bibinfo {year}
  {2012})}\BibitemShut {NoStop}%
\bibitem [{SI()}]{SI}%
  \BibitemOpen
  \href@noop {} {\ }\bibinfo {note} {See supplementary information}\BibitemShut
  {NoStop}%
\bibitem [{\citenamefont {Shi}\ \emph {et~al.}(2008)\citenamefont {Shi},
  \citenamefont {He},\ and\ \citenamefont {Anand}}]{Shi_OL_08_InP}%
  \BibitemOpen
  \bibfield  {author} {\bibinfo {author} {\bibfnamefont {Y.}~\bibnamefont
  {Shi}}, \bibinfo {author} {\bibfnamefont {S.}~\bibnamefont {He}}, \ and\
  \bibinfo {author} {\bibfnamefont {S.}~\bibnamefont {Anand}},\ }\href@noop {}
  {\bibfield  {journal} {\bibinfo  {journal} {Opt. Lett.}\ }\textbf {\bibinfo
  {volume} {33}},\ \bibinfo {pages} {1927} (\bibinfo {year}
  {2008})}\BibitemShut {NoStop}%
\bibitem [{\citenamefont {Zhang}\ \emph {et~al.}(2011)\citenamefont {Zhang},
  \citenamefont {McKnight}, \citenamefont {Engin}, \citenamefont {Watson},
  \citenamefont {Cryan}, \citenamefont {Gu}, \citenamefont {Thompson},
  \citenamefont {Calvez}, \citenamefont {O'Brien},\ and\ \citenamefont
  {Dawson}}]{JLOBrien_APL_11_GaN_coupler}%
  \BibitemOpen
  \bibfield  {author} {\bibinfo {author} {\bibfnamefont {Y.}~\bibnamefont
  {Zhang}}, \bibinfo {author} {\bibfnamefont {L.}~\bibnamefont {McKnight}},
  \bibinfo {author} {\bibfnamefont {E.}~\bibnamefont {Engin}}, \bibinfo
  {author} {\bibfnamefont {I.~M.}\ \bibnamefont {Watson}}, \bibinfo {author}
  {\bibfnamefont {M.~J.}\ \bibnamefont {Cryan}}, \bibinfo {author}
  {\bibfnamefont {E.}~\bibnamefont {Gu}}, \bibinfo {author} {\bibfnamefont
  {M.~G.}\ \bibnamefont {Thompson}}, \bibinfo {author} {\bibfnamefont
  {S.}~\bibnamefont {Calvez}}, \bibinfo {author} {\bibfnamefont {J.~L.}\
  \bibnamefont {O'Brien}}, \ and\ \bibinfo {author} {\bibfnamefont {M.~D.}\
  \bibnamefont {Dawson}},\ }\href@noop {} {\bibfield  {journal} {\bibinfo
  {journal} {Appl. Phys. Lett.}\ }\textbf {\bibinfo {volume} {99}},\ \bibinfo
  {pages} {161119} (\bibinfo {year} {2011})}\BibitemShut {NoStop}%
\bibitem [{\citenamefont {Griffel}\ \emph {et~al.}(2000)\citenamefont
  {Griffel}, \citenamefont {Abeles}, \citenamefont {Menna}, \citenamefont
  {Braun}, \citenamefont {Connolly},\ and\ \citenamefont
  {King}}]{Griffel_PTL_00_InGaAsP_coupler}%
  \BibitemOpen
  \bibfield  {author} {\bibinfo {author} {\bibfnamefont {G.}~\bibnamefont
  {Griffel}}, \bibinfo {author} {\bibfnamefont {J.~H.}\ \bibnamefont {Abeles}},
  \bibinfo {author} {\bibfnamefont {R.~J.}\ \bibnamefont {Menna}}, \bibinfo
  {author} {\bibfnamefont {A.~M.}\ \bibnamefont {Braun}}, \bibinfo {author}
  {\bibfnamefont {J.~C.}\ \bibnamefont {Connolly}}, \ and\ \bibinfo {author}
  {\bibfnamefont {M.}~\bibnamefont {King}},\ }\href@noop {} {\bibfield
  {journal} {\bibinfo  {journal} {IEEE Photon. Technol. Lett.}\ }\textbf
  {\bibinfo {volume} {12}},\ \bibinfo {pages} {146} (\bibinfo {year}
  {2000})}\BibitemShut {NoStop}%
\bibitem [{\citenamefont {Rabiei}\ \emph {et~al.}(2013)\citenamefont {Rabiei},
  \citenamefont {Ma}, \citenamefont {Khan}, \citenamefont {Chiles},\ and\
  \citenamefont {Fathpour}}]{RabieiOE13}%
  \BibitemOpen
  \bibfield  {author} {\bibinfo {author} {\bibfnamefont {P.}~\bibnamefont
  {Rabiei}}, \bibinfo {author} {\bibfnamefont {J.}~\bibnamefont {Ma}}, \bibinfo
  {author} {\bibfnamefont {S.}~\bibnamefont {Khan}}, \bibinfo {author}
  {\bibfnamefont {J.}~\bibnamefont {Chiles}}, \ and\ \bibinfo {author}
  {\bibfnamefont {S.}~\bibnamefont {Fathpour}},\ }\href@noop {} {\bibfield
  {journal} {\bibinfo  {journal} {Opt. Express}\ }\textbf {\bibinfo {volume}
  {21}},\ \bibinfo {pages} {25573} (\bibinfo {year} {2013})}\BibitemShut
  {NoStop}%
\bibitem [{\citenamefont {Gorodetsky}\ and\ \citenamefont
  {Ilchenko}(1999)}]{GorodetskyJOSAB99}%
  \BibitemOpen
  \bibfield  {author} {\bibinfo {author} {\bibfnamefont {M.~L.}\ \bibnamefont
  {Gorodetsky}}\ and\ \bibinfo {author} {\bibfnamefont {V.~S.}\ \bibnamefont
  {Ilchenko}},\ }\href@noop {} {\bibfield  {journal} {\bibinfo  {journal} {J.
  Opt. Soc. America B}\ }\textbf {\bibinfo {volume} {16}},\ \bibinfo {pages}
  {147} (\bibinfo {year} {1999})}\BibitemShut {NoStop}%
\bibitem [{\citenamefont {Cai}\ \emph {et~al.}(2000)\citenamefont {Cai},
  \citenamefont {Painter},\ and\ \citenamefont {Vahala}}]{CaiPRL00}%
  \BibitemOpen
  \bibfield  {author} {\bibinfo {author} {\bibfnamefont {M.}~\bibnamefont
  {Cai}}, \bibinfo {author} {\bibfnamefont {O.}~\bibnamefont {Painter}}, \ and\
  \bibinfo {author} {\bibfnamefont {K.~J.}\ \bibnamefont {Vahala}},\
  }\href@noop {} {\bibfield  {journal} {\bibinfo  {journal} {Phys. Rev. Lett.}\
  }\textbf {\bibinfo {volume} {85}},\ \bibinfo {pages} {74} (\bibinfo {year}
  {2000})}\BibitemShut {NoStop}%
\bibitem [{\citenamefont {Armani}\ \emph {et~al.}(2003)\citenamefont {Armani},
  \citenamefont {Kippenberg}, \citenamefont {Spillane},\ and\ \citenamefont
  {Vahala}}]{ArmaniNat03}%
  \BibitemOpen
  \bibfield  {author} {\bibinfo {author} {\bibfnamefont {D.}~\bibnamefont
  {Armani}}, \bibinfo {author} {\bibfnamefont {T.}~\bibnamefont {Kippenberg}},
  \bibinfo {author} {\bibfnamefont {S.}~\bibnamefont {Spillane}}, \ and\
  \bibinfo {author} {\bibfnamefont {K.}~\bibnamefont {Vahala}},\ }\href@noop {}
  {\bibfield  {journal} {\bibinfo  {journal} {Nature}\ }\textbf {\bibinfo
  {volume} {421}},\ \bibinfo {pages} {925} (\bibinfo {year}
  {2003})}\BibitemShut {NoStop}%
\bibitem [{\citenamefont {Sumetsky}(2004)}]{SumetskyOE04}%
  \BibitemOpen
  \bibfield  {author} {\bibinfo {author} {\bibfnamefont {M.}~\bibnamefont
  {Sumetsky}},\ }\href@noop {} {\bibfield  {journal} {\bibinfo  {journal} {Opt.
  Express}\ }\textbf {\bibinfo {volume} {12}},\ \bibinfo {pages} {2303}
  (\bibinfo {year} {2004})}\BibitemShut {NoStop}%
\bibitem [{\citenamefont {P{\"o}llinger}\ \emph {et~al.}(2009)\citenamefont
  {P{\"o}llinger}, \citenamefont {O’Shea}, \citenamefont {Warken},\ and\
  \citenamefont {Rauschenbeutel}}]{PollingerPRL09}%
  \BibitemOpen
  \bibfield  {author} {\bibinfo {author} {\bibfnamefont {M.}~\bibnamefont
  {P{\"o}llinger}}, \bibinfo {author} {\bibfnamefont {D.}~\bibnamefont
  {O’Shea}}, \bibinfo {author} {\bibfnamefont {F.}~\bibnamefont {Warken}}, \
  and\ \bibinfo {author} {\bibfnamefont {A.}~\bibnamefont {Rauschenbeutel}},\
  }\href@noop {} {\bibfield  {journal} {\bibinfo  {journal} {Phys. Rev. Lett.}\
  }\textbf {\bibinfo {volume} {103}},\ \bibinfo {pages} {053901} (\bibinfo
  {year} {2009})}\BibitemShut {NoStop}%
\bibitem [{\citenamefont {Sumetsky}\ \emph {et~al.}(2010)\citenamefont
  {Sumetsky}, \citenamefont {Dulashko},\ and\ \citenamefont
  {Windeler}}]{SumetskyOL10}%
  \BibitemOpen
  \bibfield  {author} {\bibinfo {author} {\bibfnamefont {M.}~\bibnamefont
  {Sumetsky}}, \bibinfo {author} {\bibfnamefont {Y.}~\bibnamefont {Dulashko}},
  \ and\ \bibinfo {author} {\bibfnamefont {R.}~\bibnamefont {Windeler}},\
  }\href@noop {} {\bibfield  {journal} {\bibinfo  {journal} {Opt. Lett.}\
  }\textbf {\bibinfo {volume} {35}},\ \bibinfo {pages} {898} (\bibinfo {year}
  {2010})}\BibitemShut {NoStop}%
\end{thebibliography}%

\end{document}